# Photomolecular Effect Leading to Water Evaporation Exceeding Thermal Limit

**Authors:** Yaodong Tu[1,2], Jiawei Zhou[1], Shaoting Lin[1], Mohammed AlShrah[1], Xuanhe Zhao[1]\*, and Gang Chen[1]\*

**Affiliations:**

[1]Department of Mechanical Engineering, Massachusetts Institute of Technology; Cambridge, MA 02139, USA

[2]School of Mechanical Engineering, Shanghai Jiao Tong University; Shanghai, 200240, CN

\*Corresponding authors: gchen2@mit.edu; zhaox@mit.edu

**Abstract:** We report the discovery of photomolecular effect: cleavage of water clusters off surfaces by photons. This effect is demonstrated through surprising absorption of partially wetted hydrogel in the visible spectrum where both water and hydrogel materials' absorption are negligible. Illumination of hydrogel under solar or visible-spectrum light-emitting-diode leads to evaporation rates exceed the thermal evaporation limit, even in hydrogels without additional absorbers. Measurements of vapor temperature and vapor spectrum above evaporating surfaces show clear signatures of water clusters. The photomolecular effect happens at liquid-vapor interface due to large electrical field gradients and quadrupole force on molecular clusters. This photomolecular evaporation process might be happening widely in nature, potentially impacting climate and plants growth, and can be exploited for clean water and drying technologies.

**One-Sentence Summary:** This paper shows that photons in the visible spectrum can cleave off water clusters from water-vapor interface despite weak absorption of bulk water in this spectrum, named as the photomolecular effect, which leads to evaporation from porous hydrogels under normal solar or visible light radiation exceeding the thermal evaporation limit.

**Main Text:** Water absorbs very little in the visible spectrum of sunlight; for example, the penetration length of light with 0.5 µm wavelength reaches 40 m (*1, 2*). Additional absorption materials are needed to heat up water using sunlight (*3, 4*). In the solar interfacial-evaporation approaches that have received great attention in recent years, a wide range of absorbers have been used as additives to porous host materials floating on water surfaces where sunlight are absorbed to evaporate water (*5–9*). Of particular interest is the observation that polyvinyl alcohol (PVA) hydrogel embedded with different absorbing materials can have evaporation rates exceeding the theoretical thermal evaporation limit (*9, 10*). Since the original report (*9*), different porous host materials, both organic and inorganic, and absorbers, such as electrically conducting polymers, TiO$_2$ nanoparticles, and carbon have been shown to be able to exceed the thermal evaporation limit (*9–20*). The mechanism of this phenomenon remains unclear, although reduced latent heat of water in hydrogels has been hypothesized (*9, 10*). In this paper, we discover a photomolecular effect which makes water absorbing in the spectrum range where bulk water is least absorbing. In this effect, water clusters are cleaved off directly by photons at the water-vapor interface. This mechanism bears similarity to the photoelectric effect for electron emission discovered by Hertz (*21*) and explained by Einstein (*22*), but also with three significant differences (i) no electronic transition is involved, (ii) it happens in the spectrum where bulk water does not absorb and (iii) one photon can cleave off a cluster of water molecules. The cleaved water molecules can recondense (internal photomolecular effect) or escape (external photomolecular effect) under proper conditions. The escape of cleaved water molecules could lead to evaporation rate exceeding the thermal limit as we demonstrate in PVA hydrogels with and without additional absorbers under illumination of solar radiation or light-emitting-diodes (LEDs). Such photomolecular evaporation process could be happening widely in nature, from clouds to fogs, from ocean to soil surfaces, and in plant transpiration, and may have significant impacts on earth's water cycle and climate change, as well as potential applications in clean-water and drying technologies.

The theoretical limit for thermal evaporation, using solar or LED energy or electrical sources, can be calculated from $J_{t,max} = q/[L + c_p(T_s - T_w)]$, where $q$ is the solar flux or other forms of input power, $L$ the latent heat, $c_p$ the constant-pressure specific heat, $T_s$ and $T_w$ are the evaporating surface and bulk water temperatures, respectively. Taking $T_s = 40°C$ and $T_w = 20°C$, and standard solar flux at one sun $q = 1$ kW/m$^2$, the thermal evaporation limit is $\dot{m} = 1.45$ kg/(m$^2$h), assuming all solar flux converts into heat used for water evaporation. Experimentally, $J_{max}$ as high as $4 - 5 \text{ kg/(m}^2\text{h})$ (*17*, *20*) and over $10 \text{ kg/(m}^2\text{h})$ (*12*) in two- and three-dimensional structures had been reported, respectively. Different materials, starting from PVA (*9*), to other polymers (*16*, *18*, *20*) and even inorganic porous absorbers (*12*, *15*, *19*), have shown this effect. All of these works either impregnate absorbers such as conducting polymers or light-absorbing nanoparticles into porous structures, or directly use porous absorbers such as porous carbons obtained from carbonizing plants. The most-cited mechanism for the higher evaporation rate than the theoretical thermal limit is reduced latent heat of water in these materials (*9*, *10*, *14*). This mechanism was inferred based on the following reasons (*9*, *14*): (i) differential scanning calorimetry (DSC) measurement had shown latent heat reduction, (ii) in dark conditions, the evaporation rates from samples are larger than from pure water, and (iii) a theoretical picture of different water states inside hydrogel: bond water, intermediate water, and bulk water (*23*, *24*). However, DSC measurement only showed up to 30% latent heat reduction integrated over a wide temperature range (*9*, *10*), and there is no solid foundation why the intermediate water can have large reduced latent heat comparable to experimental observations. In addition, the large amount of bulk water with no latent-heat reduction coexisting in hydrogel also needs to be evaporated. The higher evaporation rates in dark conditions can be explained by the porous structures that increase the evaporating surface area. The possibility of water evaporating as clusters was mentioned in the original publication (*9*) but there was no mechanism to support such a possibility. We show below the photomolecular effect in which photons can cleave off water clusters directly without going through thermal processes. This mechanism leads to water evaporation rate under solar and narrow spectrum LED radiation exceeding the theoretical thermal evaporation limits.

We synthesized three types of porous polyvinyl alcohol (PVA) hydrogel samples (*25*): (1) pure PVA samples that do not include any additional absorbers, which are denoted as pure-PVA, (2) PVA samples integrated with polypyrrole (ppy) denoted as PVA-ppy, and (3) pure PVA coated on porous carbon paper, denoted as PVA-carbon. The synthesis involves freeze-thawing or freeze-drying to form proper porous structures (Figs. 1A-B, and Fig. S1). More structural and thermal properties of the samples are provided in supplementary materials (*25*) (see Fig. S1-S3).

We extract absorptance of different samples from measured reflectance and transmittance using an integrating sphere, which collects scattered light in addition to direct transmission and reflection (*25*) [Fig. 1C, Fig. S4]. First, we show the absorptance of pure water, dry PVA powder, the solution before forming the gel, and the as-gelated sample in Fig. 1D and Fig. S5. Above 500 nm and up to 800 nm, none of these samples show much absorption (the non-zero absorption less than 2.5% are due to experimental uncertainties), consistent with expectations due to the weak absorption of water, PVA, and cross-linkers and initiators used in the sample preparation. Surprisingly, although the freeze-thawed samples (Fig. 1D, Fig.S6) still have about the same amount of water as the as-gelated sample, its absorptance increased significantly. In Figs. 1E&F and Fig. S7, we show absorptance of freeze-dried pure-PVA samples with different amount of water added back to control the water content in the samples. Although dry pure-PVA does not absorb, the samples become absorbing as some waters is added back, and the absorptance eventually decreases with more water. Considering that multiple scattering effect can increase absorption pathlength by a maximum of $4n^2$ (*26*), where $n$ is the refractive index (~1.3 for both water and polymers), the large absorption cannot possibly be explained by the light trapping effect.

The surprising absorption in the freeze-thawed and freeze-dried pure-PVA samples containing some water can be explained by invoking a photomolecular effect which we will describe here and support with more experiments later. Inside bulk water, very little absorption exists in the visible spectrum [Fig. S5] because the photon energies are too high for the intramolecular vibrational modes (some residual absorption in water in the visible range (*1*) was thought as due to higher harmonics of intramolecular vibrational modes (*27*)) and even longer wavelength intermolecular vibrational and librational modes (*28*), and the photon energies are too low for the electronic transition in the ultraviolet region. In water, hydrogen bond dominates, with typical bonding energy 0.22 – 0.26 eV (*29*). It is well-known that water molecules form fluctuating clusters due to the hydrogen bond, although the exact pictures, such as the size and shape and lifetime of the clusters, are still under debate (*29–33*). Theoretically, a photon can cleave several water molecules together as a cluster by breaking bonds between the cluster and rest of water. Taking a photon at 500 nm with an energy 2.48 eV, it can cleave off ~10 or even more intermolecular water bonds, depending on if these are hydrogen bonds or even weaker van der Waals bond. In bulk water, however, there is no space for the cluster to escape, i.e., the final states are occupied, and hence the process is forbidden. On the other hand, this process can happen at the surface or internal voids in the liquids. The surface layer thickness of water is around 3 – 7 Å (*34, 35*). Such a thin region is not enough to cause appreciable absorption in bulk water with a flat interface. In freeze-dried pure-PVA samples with controlled water contents, there are increased water-vapor interface areas for photons to directly cleave off water clusters, which enter air, leading to measurable absorption (Figs.1E&F, Fig. S7). In freeze-thawed samples, we hypothesize polymer conformational change creates internal voids that allow the water clusters to be cleaved off and recondense [Fig. S6]. The former is the external photomolecular effect and the latter is the internal photomolecular effect. Next, we will show more experimental evidence supporting the photomolecular mechanism, followed by further discussion of the physical picture.

In Figs. 2A and Fig. S8, we show the evaporation measurement set up (*25*). Figures 2B&C are the typical evaporation history of hydrogel samples under solar radiation. The evaporation has two stages. In the initial stage, the evaporation rate is lower, below the thermal limit. We observe that in this stage, the surface of the sample under testing is still flooded with water. This is the normal thermal evaporation stage and the evaporation rate never exceeds the thermal limit. The second evaporation stage commences when the water surface recesses below the sample top surface. In this stage, the measured evaporation rate of both PVA-ppy and PVA-carbon samples exceed the thermal evaporation limit. Although the measured evaporation rates of pure-PVA samples are below the limit due to its low absorptance (Fig. 1E&F, Fig. S7) compared to PVA-ppy and PVA-carbon samples (Fig.S9), the evaporate rates normalized to the measured absorptance are even higher than PVA-ppy and PVA-carbon samples (Fig. 2D). Although we do not know the exact absorbance of pure PVA sample in operation, the samples did not seem change much visually during evaporation, and measured absorptance of pure-PVA samples do not change much in a wide range of water content as seen in Fig. 1F. Hence, we took an absorptance value of 20% in the normalization, which is the maximum of measured absorptance and hence is a conservative value. In Fig. 2C, we also show evaporation from PVA-carbon samples when the solar radiation is shined from the back side of the sample (*25*) (Fig. S10). We adjust the solar radiation such that the surface temperature of the sample is the same as irradiation from the front side. The experiment does not show two stages despite that water recess below surface, and evaporation is pure thermal, with rate below the thermal limit.

We used LED lamps of different wavelengths to carry out evaporation tests (*25*) and found that the evaporation rates in stage 2 depends on wavelengths. Figures 2E and 2F show evaporation rates in this stage at different wavelengths for both PVA-ppy and pure-PVA samples, both of which exhibit a peak evaporation rate at 520 nm wavelength. Note that the absorptance itself does not show peaks in the visible spectrum (Fig. 1D). Also plotted in these figures are the temperature of the sample surface, measured using thermocouples (*25*). Chen et al. (*36*) reported different evaporation rate of MnO$_2$ nanowire on polystyrene foam under UV, Vis, and IR spectra of sunlight. Although they did not give exact spectrum nor report details how data were normalized, they did show highest evaporation rate from visible light.

We interpret the peak evaporation rate at 520 nm wavelength as due to the size of evaporated water clusters is probably maximum at this wavelength. Lower evaporation rate under shorter wavelength lights could be due to the competition between the photomolecular and photothermal effects, while lower evaporation rates at longer wavelengths could be due to smaller clusters that can be excited by one photon. Further increase in wavelength leads to photothermal heating due to bulk water absorption. The surface temperature monotonically increases with wavelength, which we explain as due to the smaller penetration depths of short wavelength photons, leading to evaporation closer to surface and easier vapor escape from the surface. Molecular clusters cleaved by longer wavelength lights beneath the surface can re-condense (the internal photomolecular mechanism) and release heat in the process, hence showing higher temperatures.

We also tested evaporation rate under different solar intensities. Interestingly, at lower light intensities, the evaporation rate normalized to the input energy is higher than that at higher light intensities (Fig. 2G and 2H), consistent with the picture that the evaporation is due to both photomolecular and photothermal processes. We anticipate that the photomolecular mechanism has no or weak temperature dependence, while it is well-known that thermal evaporation increases strongly with temperature. The measured trend is consistent with this anticipation: the sample temperature rise is smaller at lower light intensity. In fact, evaporation from the pure-PVA sample under 0.1 sun green LED leads to a surface temperature of 21.5 ºC, lower than the ambient temperature of 22.4 ºC (Fig. 2G).

We also tested purely thermal evaporation by embedded electrical heaters inside sample (*25*) (Fig. S10). The evaporation rate never exceeds thermal limits. Figure 2I compares the weight loss and surface temperature of the same sample under solar radiation and joule heating, keeping the same surface temperature. We note that solar heating reaches the steady state much faster than joule heating, in addition to a faster evaporation rate, again demonstrating the difference between photomolecular evaporation and thermal evaporation.

If water clusters are cleaved off, we expect to see signatures in the vapor layer above the evaporating samples. Here we show two important signatures: temperature distributions and transmittance spectra at different heights above the surface. First, Fig. 3A shows temperature distribution in the vapor phase when the light is on and immediately after light is blocked off for a PVA-ppy sample, measured using an infrared camera for thermal radiation from a thin glass slide suspended in the vapor phase as the thermal emitter (*25*) (Fig. S11). We notice that when the light is on so that the photomolecular effect is at play, the vapor temperature drop within the first 2 mm above the sample surface is much faster than when the light is off but the sample surface temperatures almost don't change (~35.4 °C), i.e., thermal evaporation, for which the temperature distribution drops almost linearly away from the surface. Figure 3B compares the temperature distribution above surface between solar heating and joule heating, measured using a specially shaped thermocouple (*25*) (Fig. S12). Under solar irradiation, the temperature peaks near surface, and drops sharply, consistent with the infrared image. Under electrical heating, the temperature distribution is similar to that after light is blocked off as shown in Fig. 3A. Interestingly, the temperature distribution between 6 – 13 mm above the samples are almost constant under solar irradiation. Such behavior is also seen in the vapor region above pure-PVA sample, while evaporation from pure water surface at the similar evaporation surface temperature (~35.8 °C) achieved by placing an absorber at the bottom of the container and adjusting the sunlight intensity (Fig. 3C) shows behavior similar to normal thermal evaporation.

We interpret the sharp drop of temperature near surface region as due to disintegration of water clusters when they collide with air molecules, absorbing heat which leads to a sharp temperature drop (Fig. 3D), and the flat temperature region is due to supersaturation after the clusters breakup. In this flat region, clusters break and re-nucleate as air becomes supersaturated. In fact, we can visibly see condensation on a glass slide under green light (1 sun) over a PVA-ppy sample (Video 1), even its evaporation temperature is only about 42 °C. The brightening and dimming of the images are interpreted as due to emittance change caused by interference effect as the condensed liquid layer thickness changes.

Figure 3E shows transmission spectrum of light in the vapor phase measured with beam at different height above the evaporating surface from different samples and Fig. 3F compares the spectra of different samples at the same heights. The testing setup is shown in Fig. S13 (*25*). The transmittance is normalized to that of measured spectrum when the sample chamber is filled with dry nitrogen. The spectra are rich and we do not expect to be able to explain all details. We note the key features: for pure water, the main peaks do not change with heights. However, we note vapor transmission spectrum above pure-PVA and PVA-ppy samples are generally quite different from vapors evaporating from pure water surface in 0 – 3 mm range from the surface. We note there are regions the absorption position blue shift (as marked down in Fig. 3E) as we move away from surface. This is because of the strength of intramolecular vibration as clusters dissociate and become smaller and less affected by hydrogen bonds with other molecules. There are also regions the absorption positions red shift, which could be due to the change of librational modes of the clusters. Water clusters are difficult to create and to measure. In the past, supersonic expansion of jet and helium bubbling techniques had been used to create water clusters and different spectroscopic techniques were employed to study the clusters (*37, 38*). The fact that we can directly observe the vapor spectra change above evaporating sample surfaces indicates the abundance of the clusters, although the complexity of the spectra prevents us from identifying the cluster size and distributions. More data on the transmission spectra of light in the vapor phase are shown in Fig. S14 and data files are provided.

We believe the above experimental evidence have unequivocally shown the photomolecular effect: direct photon cleavage of water clusters from water-vapor interface. We admit that details of the photon interaction with water at the interface remain to be clarified. This is not surprising considering the long history (*39*) in developing quantitative theories for photoelectric effect since Hertz' discovery and Einstein's qualitative picture. For example, to explain the surface photoelectric effect, one needs to include the failure of Maxwell equations and consider nonlocal dielectric constant (*40, 41*). This can be understood since one of the boundary conditions for the Maxwell equations is the continuity of the perpendicular component of the displacement field, which leads to $\varepsilon_1 E_{1\perp} = \varepsilon_2 E_{2\perp}$, where $\varepsilon$ is the dielectric constants and subscripts 1 and 2 represent quantities on the two sides and $E_\perp$ is the electrical field perpendicular to the interface. This equation implies a discontinuity of the electric field, which is, of course, a mathematical simplification. In reality, the field changes rapidly from one side to the other. At a metal-dielectric interface, electron wavefunction spill out to the dielectric side, and its density changes rapidly from the dielectric side to metal bulk values, over a distance of the order of a few Å. To model photoemission current, both the rapid electric field change and the electron density change need to be considered, including nonlocal dielectric constants (*40, 41*).

The above picture gives some clues to the driving force of photomolecular effect as shown in Fig. 4a. It is known that molecular density at water surface changes rapidly to its vapor density over a distance around 3 – 7 Å (*34, 35*). Correspondingly, the perpendicular component of the electric field will also change from its value in the vapor phase to that inside water over a comparable distance, leading to a large gradient of the electrical field. Of course, a significant difference of a liquid water-vapor interface from that of a metal-dielectric interface is that the water molecules themselves are neutral. However, water is polar and the single water dipole moment is around 1.8D and increases to ~2.8D for water clusters (*37*) (where D denotes Debye dipole moment unit), suggesting an effective charge separation ~0.5 Å. Such charge separation under the large electric field gradient at the interface leads to a net force on the molecule, i.e., the quadrupole force (*42*). When the force points outward during the cycle of the time-varying field, the water clusters can be driven off the liquid surface. The difference of the photon energy $\hbar\omega$ to the bonding energy $\Delta$ of a water cluster to the surrounding water $\hbar\omega - \Delta$ is converted into the kinetic energy of the molecular cluster (Fig.4b) (a fraction can also dissipate inside the liquid during bond breaking). The bonding energy $\Delta$ consists of multiple bonds of the cluster with the surrounding, $\Delta \sim n\Delta_1$, where $n$ is the number of bonds and $\Delta_1$ is the average energy of one bond. At this stage, we do not know if all bonds are of identical energy, which we think it is unlikely, nor do we know if all of them are non-hydrogen bonds. This molecular process can only happen at a liquid-vapor interface.

After a molecular cluster is cleaved off the surface, the clusters will collide with other vapor and air molecules, which could change the directions and/or break up the molecules in the cluster (Fig. 4c). However, the breaking up events are infrequent since the average thermal energy is $k_B T \sim 0.026$ eV at room temperature, while the hydrogen bonds between molecules in the cluster is $E_1 \sim 0.22 - 0.26$ eV. The lifetime $\tau_c$ for breaking up one molecule is $\exp\left(\frac{E_1}{k_B T}\right) \sim 10^4$ times longer than the regular molecular collision time $\tau$ of the order of ns. The clusters may change their direction due to momentum exchange with air molecules during each collision. We can estimate the average distance between breaking up one molecule from the cluster as $\Lambda\sqrt{\tau_c/\tau}$, where $\Lambda$ is the mean free path between collision, which is ~100 nm for air molecules. This means the breaking up will happen between 10 – 1000 μm, considering that the molecule clusters have multiple molecules and also may have different initial velocities. Heat is absorbed in this region, which explains why we measured a sharp temperature drop in this region (Region I). After the vapor becomes supersaturated, the breakup and recondensation of water molecules can simultaneously happen, explaining the flat region we observe in Figs. 3A-C (Region II). If the clusters are cleaved off deep inside the hydrogel, they can also re-condense and release heat. This process belongs to the internal photomolecular effect that we observe absorption in freeze-thawed samples (Fig. 1D).

There is the possibility that light can also dissociate the molecular clusters as is known in the blackbody infrared radiative dissociation phenomenon (*43*). Photon flux from both the solar radiation and environmental blackbody radiation at 300 K are at 0.26 eV are both ~$10^{20}$ $m^{-2}s^{-1}\mu m^{-1}$, suggesting that this process could also be involved.

In summary, we have demonstrated the existence of the photomolecular effect: photons cleave off water clusters from the surface region. The photomolecular evaporation can be internal and external, although our study here emphasizes external effect which can lead to increased evaporation rate above the thermal evaporation limit. Although we provide qualitative explanations for why the photomolecular effect happens, there are certainly a lot more work remain to be done in further characterizing the photomolecular effect, understanding the mechanisms, and applying as a tool for fundamental science and developing useful technologies. Our observations lead us to question if this effect happens widely in nature such as in clouds, fogs, ocean and soil surfaces, plant transpiration, and to other liquids. Answering these questions calls for collaborations from researchers from different fields.

**References and Notes**

1. G. M. Hale, M. R. Querry, Optical constants of water in the 200-nm to 200-μm wavelength region. *Appl. Opt.* **12**, 555–563 (1973).
2. T. A. Cooper, S. H. Zandavi, G. W. Ni, Y. Tsurimaki, Y. Huang, S. V. Boriskina, G. Chen, Contactless steam generation and superheating under one sun illumination. *Nat. Commun.* **9**, 1–10 (2018).
3. T. P. Otanicar, P. E. Phelan, R. S. Prasher, G. Rosengarten, R. A. Taylor, Nanofluid-based direct absorption solar collector. *J. Renew. Sustain. Energy*. **2**, 033102 (2010).
4. O. Neumann, A. S. Urban, J. Day, S. Lal, P. Nordlander, N. J. Halas, Solar vapor generation enabled by nanoparticles. *ACS Nano*. **7**, 42–49 (2012).
5. H. Ghasemi, G. Ni, A. M. Marconnet, J. Loomis, S. Yerci, N. Miljkovic, G. Chen, Solar steam generation by heat localization. *Nat. Commun.* **5**, 1–7 (2014).
6. Z. Wang, Y. Liu, P. Tao, Q. Shen, N. Yi, F. Zhang, Q. Liu, C. Song, D. Zhang, W. Shang, T. Deng, Bio-inspired evaporation through plasmonic film of nanoparticles at the air–water interface. *Small*. **10**, 3234–3239 (2014).
7. P. Tao, G. Ni, C. Song, W. Shang, J. Wu, J. Zhu, G. Chen, T. Deng, Solar-driven interfacial evaporation. *Nat. Energy*. **3**, 1031–1041 (2018).
8. L. Zhou, Y. Tan, J. Wang, W. Xu, Y. Yuan, W. Cai, S. Zhu, J. Zhu, 3D self-assembly of aluminium nanoparticles for plasmon-enhanced solar desalination. *Nat. Photonics*. **10**, 393–398 (2016).
9. F. Zhao, X. Zhou, Y. Shi, X. Qian, M. Alexander, X. Zhao, S. Mendez, R. Yang, L. Qu, G. Yu, Highly efficient solar vapour generation via hierarchically nanostructured gels. *Nat. Nanotechnol.* **13**, 489–495 (2018).
10. F. Zhao, Y. Guo, X. Zhou, W. Shi, G. Yu, Materials for solar-powered water evaporation. *Nat. Rev. Mater.* **5**, 388–401 (2020).
11. Z. Dong, C. Zhang, H. Peng, J. Gong, Q. Zhao, Modular design of solar-thermal nanofluidics for advanced desalination membranes. *J. Mater. Chem. A*. **8**, 24493–24500 (2020).
12. J. Li, X. Wang, Z. Lin, N. Xu, X. Li, J. Liang, W. Zhao, R. Lin, B. Zhu, G. Liu, L. Zhou, S. Zhu, J. Zhu, Over 10 kg m$^{-2}$ h$^{-1}$ Evaporation Rate Enabled by a 3D Interconnected Porous Carbon Foam. *Joule*. **4**, 928–937 (2020).
13. X. Mu, Y. Gu, P. Wang, J. Shi, A. Wei, Y. Tian, J. Zhou, Y. Chen, J. Zhang, Z. Sun, J. Liu, B. Peng, L. Miao, Energy matching for boosting water evaporation in direct solar steam generation. *Sol. RRL*. **4**, 2000341 (2020).
14. X. Zhou, F. Zhao, Y. Guo, Y. Zhang, G. Yu, A hydrogel-based antifouling solar evaporator for highly efficient water desalination. *Energy Environ. Sci.* **11**, 1985–1992 (2018).
15. Y. Bian, Q. Du, K. Tang, Y. Shen, L. Hao, D. Zhou, X. Wang, Z. Xu, H. Zhang, L. Zhao, S. Zhu, J. Ye, H. Lu, Y. Yang, R. Zhang, Y. Zheng, S. Gu, Carbonized bamboos as excellent 3D solar vapor-generation devices. *Adv. Mater. Technol.* **4**, 1800593 (2019).
16. H. Wang, R. Zhang, D. Yuan, S. Xu, L. Wang, Gas foaming guided fabrication of 3D porous plasmonic nanoplatform with broadband absorption, tunable shape, excellent stability, and high photothermal efficiency for solar water purification. *Adv. Funct. Mater.* **30**, 2003995 (2020).

17. Y. Guo, X. Zhao, F. Zhao, Z. Jiao, X. Zhou, G. Yu, Tailoring surface wetting states for ultrafast solar-driven water evaporation. *Energy Environ. Sci.* **13**, 2087–2095 (2020).

18. T. Gao, X. Wu, Y. Wang, G. Owens, H. Xu, A hollow and compressible 3D photothermal evaporator for highly efficient solar steam generation without energy loss. *Sol. RRL.* **5**, 2100053 (2021).

19. Y. Shi, C. Zhang, Y. Wang, Y. Cui, Q. Wang, G. Liu, S. Gao, Y. Yuan, Plasmonic silver nanoparticles embedded in flexible three-dimensional carbonized melamine foam with enhanced solar-driven water evaporation. *Desalination.* **507**, 115038 (2021).

20. X. Zhang, Y. Peng, L. Shi, R. Ran, Highly efficient solar evaporator based on a hydrophobic association hydrogel. *ACS Sustain. Chem. Eng.* **8**, 18114–18125 (2020).

21. H. Hertz, On an effect of ultraviolet light upon the electric discharge. *Ann. Phys.* **31**, 983 (1887).

22. A. Einstein, On a heuristic point of view concerning the production and transformation of light. *Ann. Phys.* **17**, 132–148 (1905).

23. M. S. Jhon, J. D. Andrade, Water and hydrogels. *J. Biomed. Mater. Res.* **7**, 509–522 (1973).

24. T. Ikeda-Fukazawa, N. Ikeda, M. Tabata, M. Hattori, M. Aizawa, S. Yunoki, Y. Sekine, Effects of crosslinker density on the polymer network structure in poly-N,N-dimethylacrylamide hydrogels. *J. Polym. Sci. Part B Polym. Phys.* **51**, 1017–1027 (2013).

25. Materials and methods are available as supplementary materials.

26. E. Yablonovitch, Statistical ray optics. *JOSA, Vol. 72, Issue 7, pp. 899-907*. **72**, 899–907 (1982).

27. C. L. Braun, S. N. Smirnov, Why is water blue? *J. Chem. Educ.* **70**, 612–614 (1993).

28. M. Grechko, T. Hasegawa, F. D'Angelo, H. Ito, D. Turchinovich, Y. Nagata, M. Bonn, Coupling between intra- and intermolecular motions in liquid water revealed by two-dimensional terahertz-infrared-visible spectroscopy. *Nat. Commun.* **9**, 1–8 (2018).

29. F. H. Stillinger, Water revisited. *Science (80-. ).* **209**, 451–457 (1980).

30. R. Ludwig, Water: From clusters to the bulk. *Angew. Chemie Int. Ed.* **40**, 1808–1827 (2001).

31. S. W. Benson, E. D. Siebert, A simple two-structure model for liquid water. *J. Am. Chem. Soc.* **114**, 4269–4276 (2002).

32. M. F. Chaplin, A proposal for the structuring of water. *Biophys. Chem.* **83**, 211–221 (2000).

33. F. Weinhold, Quantum cluster equilibrium theory of liquids: General theory and computer implementation. *J. Chem. Phys.* **109**, 367 (1998).

34. J. Alejandre, D. J. Tildesley, G. A. Chapela, Molecular dynamics simulation of the orthobaric densities and surface tension of water. *J. Chem. Phys.* **102**, 4574 (1998).

35. L. X. Dang, T.-M. Chang, Molecular dynamics study of water clusters, liquid, and liquid–vapor interface of water with many-body potentials. *J. Chem. Phys.* **106**, 8149 (1998).

36. B. Chen, X. Zhang, Y. Xia, G. Liu, H. Sang, Y. Liu, J. Yuan, J. Liu, C. Ma, Y. Liang, M.

Song, G. Jiang, Harnessing synchronous photothermal and photocatalytic effects of cryptomelane-type $MnO_2$ nanowires towards clean water production. *J. Mater. Chem. A*. **9**, 2414–2420 (2021).

37. J. K. Gregory, D. C. Clary, K. Liu, M. G. Brown, R. J. Saykally, The water dipole moment in water clusters. *Science (80-. )*. **275**, 814–817 (1997).

38. F. Huisken, M. Kaloudis, A. Kulcke, Infrared spectroscopy of small size-selected water clusters. *J. Chem. Phys.* **104**, 17 (1998).

39. S. Hüfner, *Photoelectron Spectroscopy : Principles and Applications* (Springer Berlin Heidelberg, 2003).

40. P. J. Feibelman, Surface electromagnetic fields. *Prog. Surf. Sci.* **12**, 287–407 (1982).

41. A. Liebsch, *Electronic Excitations at Metal Surfaces* (Springer US, 1997).

42. J. D. Jackson, *Classical Electrodynamics* (Wiley, ed. 3rd, 1998).

43. R. C. Dunbar, BIRD (blackbody infrared radiative dissociation): Evolution, principles, and applications. *Mass Spectrom. Rev.* **23**, 127–158 (2004).

44. G. Fang, C. A. Ward, Temperature measured close to the interface of an evaporating liquid. *Phys. Rev. E*. **59**, 417 (1999).

**Acknowledgments:** We would like to thank Ms. Hongxia Zeng who contributed to sample fabrication in 1/2019-10/2019 and thank Prof. Guihua Yu and his group members Dr. Fei Zhao, Xingyi Zhou and Youhong Guo for demonstrating PVA-PPy sample fabrication in 2019. We thank the following people for discussion and help during the course of this work: Dr. Jungwoo Shin (fabricated a home-made freeze-dryer with help of YDT and MA), Dr. Yoichiro Tsurimaki (trained YDT to use the UV-VIS spectrometer), Dr. Yi Huang (helped YDT use the UV-VIS spectrometer), Dr. Xiaoyu Chen (did sample surface treatment), Dr. Xin Qian (helped YDT make PVA-carbon sample), Qichen Song, Dr. Zhiwei Ding, and Buxuan Li. We would like to thank Prof. Evelyn Wang and her group members Lenan Zhang, Geoffrey Vaartstra, Carlos Marin and Chad Wilson for discussion. All of YDT's work was done at MIT, his first year post-doc fellowship was supported by Shanghai Jiao Tong University. MA was partially supported by a postdoctoral fellowship from the Natural Sciences and Engineering Research Council of Canada (NSERC). G.C. specially thank MIT for its support.

**Author contributions:** YDT carried out the majority of experiments throughout the whole duration of the project commenced in 1/2019. JWZ participated in the research in 1/2019-7/2020 via sample fabrication and characterization, including observing the wavelength dependence of LEDs driven PVA-PPy hydrogel evaporation, measuring temperature profile in the vapor region and observing flat temperature region. STL contributed to sample fabrication and characterization throughout whole duration of the work. MA contributed to the work through making different samples from PVA starting 04/2020 and contributed to data collection and sample characterization for data reported in the paper. XHZ supervised the research during GC's absence in 1/2021-6/2021 and thereafter co-supervised the hydrogel evaporation experiments. He guided the fabrication of all PVA samples, suggested and supervised details of the pure PVA evaporation experiments and the backside optical heating experiments to validate the photon-induced evaporation, and contributed to data interpretation. GC initiated and supervised the research except an absence from middle January 2021 to the end of June 2021. He organized sample fabrication effort and suggested early experiments on wavelength and intensity dependence, electrical heating, IR image in vapor phase and thermocouple mapping to validate the concept of photon-induced evaporation and led data interpretation, and supervised details of the experiments. In July 2021, he conceived the theoretical picture of photomolecular effect, suggested experiments (hydrogel absorptance and vapor spectra) to validate the photomolecular concept and led data interpretation, and supervised details of the experiments. GC led the writing of the manuscript, working closely with YDT. All team members contributed to manuscripts via discussion, figure making, and proofreading.

**Competing interests:** Authors declare that they have no competing interests.

**Data and materials availability:** All data are available in the main text or the supplementary materials.

**Supplementary Materials**

Materials and Methods

Figs. S1 to S14

Tables S1 to S3

Reference (*44*)

Movies S1

Data S1 to S4

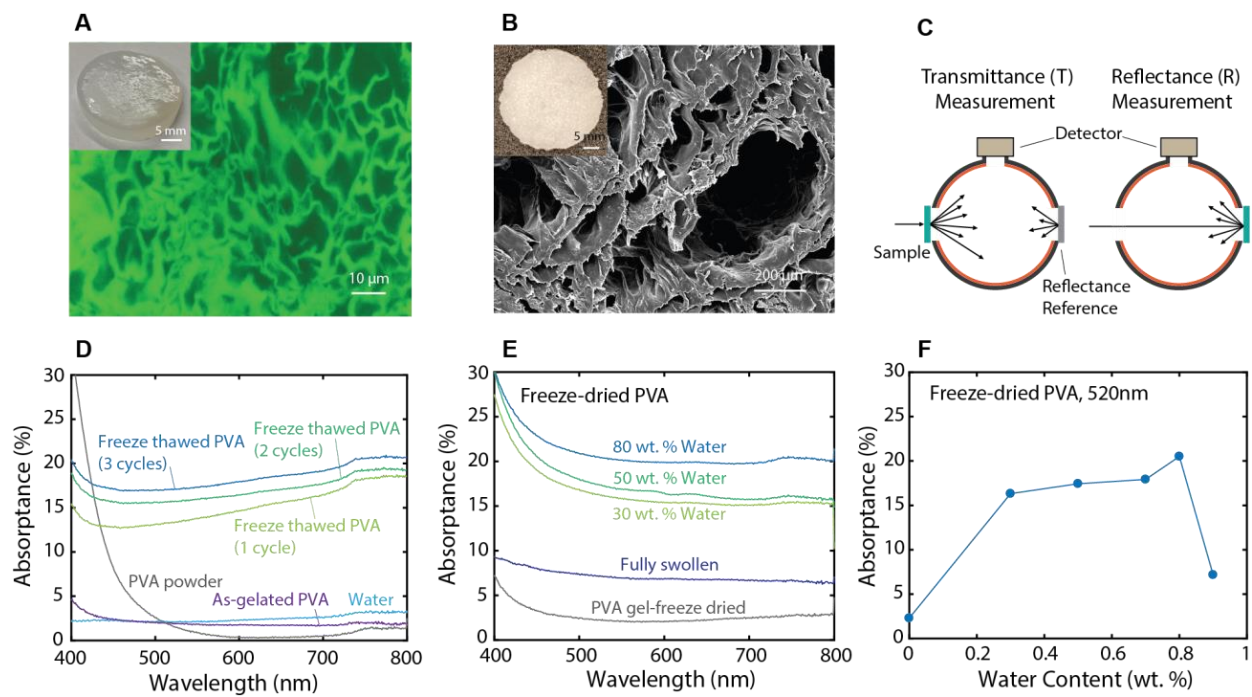

**Fig. 1. Visible light absorption of hydrogels with different water contents.** (**A**) Photo and confocal microscope image of a wet pure-PVA sample. (**B**) Photo and SEM image of a freeze-dried PVA-ppy sample. (**C**) Illustration of integrating sphere measurements of absorptance by measuring reflectance and transmittance. (**D** and **E**) Measured absorptance of different samples. Although pure water, pure PVA solution and as-gelated PVA samples do not absorb in the visible range, the freeze-thawed and freeze-dried pure-PVA samples have significant absorptance due to photomolecular effect. (**F**) Absorptance of freeze-dried pure-PVA samples as a function of the water content at 520 nm, showing that absorptance depends on water contents.

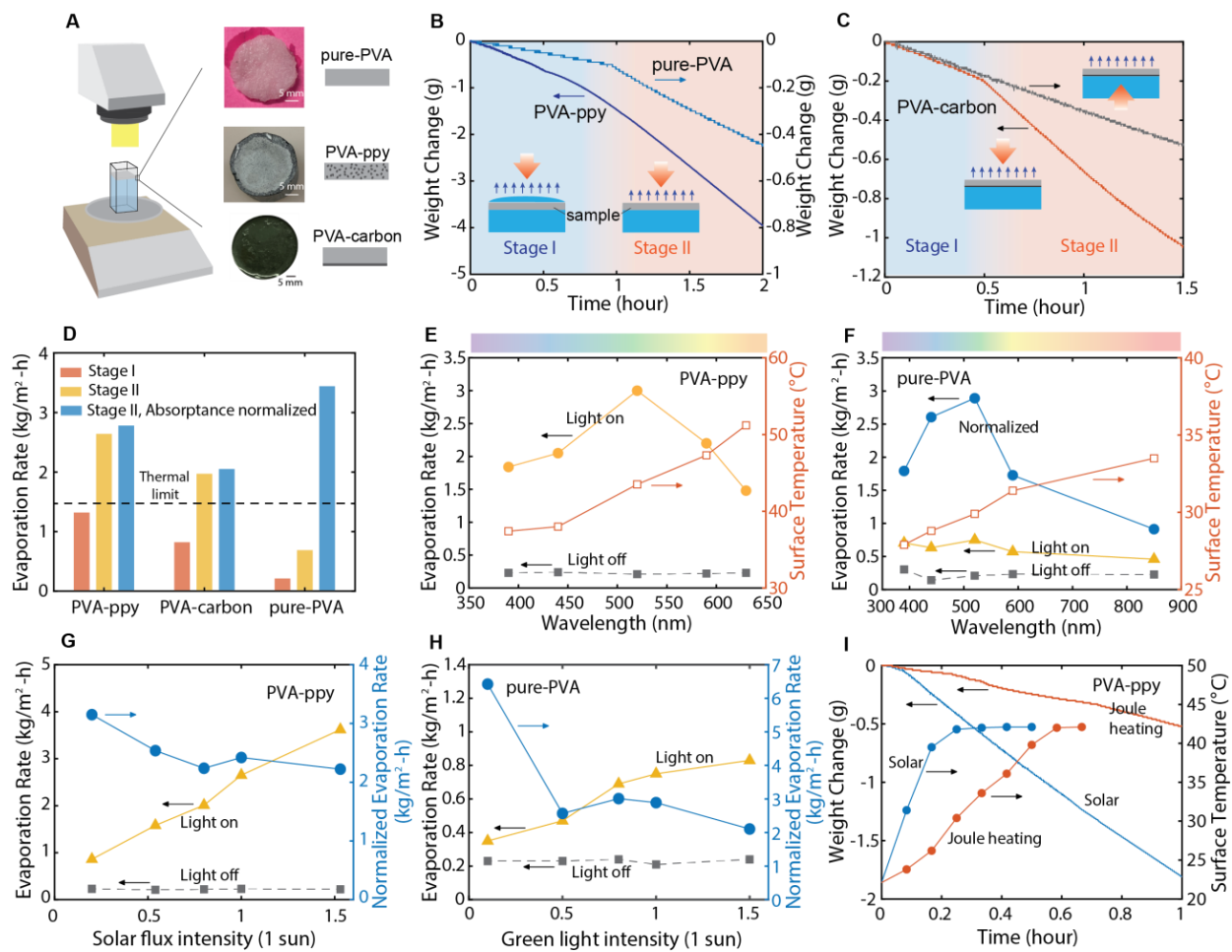

**Fig. 2. Peculiarities of light-driven hydrogel evaporation.** (**A**) Evaporation rate measurement platform and sample photos. (**B**) Weight change as a function time for a pure-PVA and a PVA-ppy sample under one sun, showing two stages. In stage I, water floods sample surface and the evaporation rate is lower than the thermal limit. This is the thermal evaporation stage. In stage II, water recesses into pores and evaporation rate increases significantly, exceeding the thermal limit. This stage has both photomolecular and thermal evaporation. (**C**) Weight change as a function of time for PVA-carbon under one sun irradiation from front and from back. Front illumination shows similar two stages as in (B), while back illumination only shows thermal evaporation. (**D**) Comparison of evaporation rates under one sun in stage I and stage II among different samples, clearly showing stage II evaporation rates exceed the thermal limit. (**E**) and (**F**) Evaporate rates variation with wavelengths using LED radiation under one-sun equivalent intensity for PVA-ppy (E) and pure-PVA samples (F), both showing a peak rate at 520 nm. Surface temperatures are also shown. (**G**) Measured evaporation rates of a PVA-ppy sample from solar simulator under different solar intensities, showing higher normalized rates at lower intensity. (**H**) Measured evaporation rate of a pure-PVA sample under green (520 nm) LED illumination, again showing higher rates at lower intensities. (**I**) Weight change and surface temperature as a function of time for a PVA-ppy sample under solar irradiation and electrical heating with the same evaporation temperatures. It takes much longer for joule heating to reach steady surface temperature.

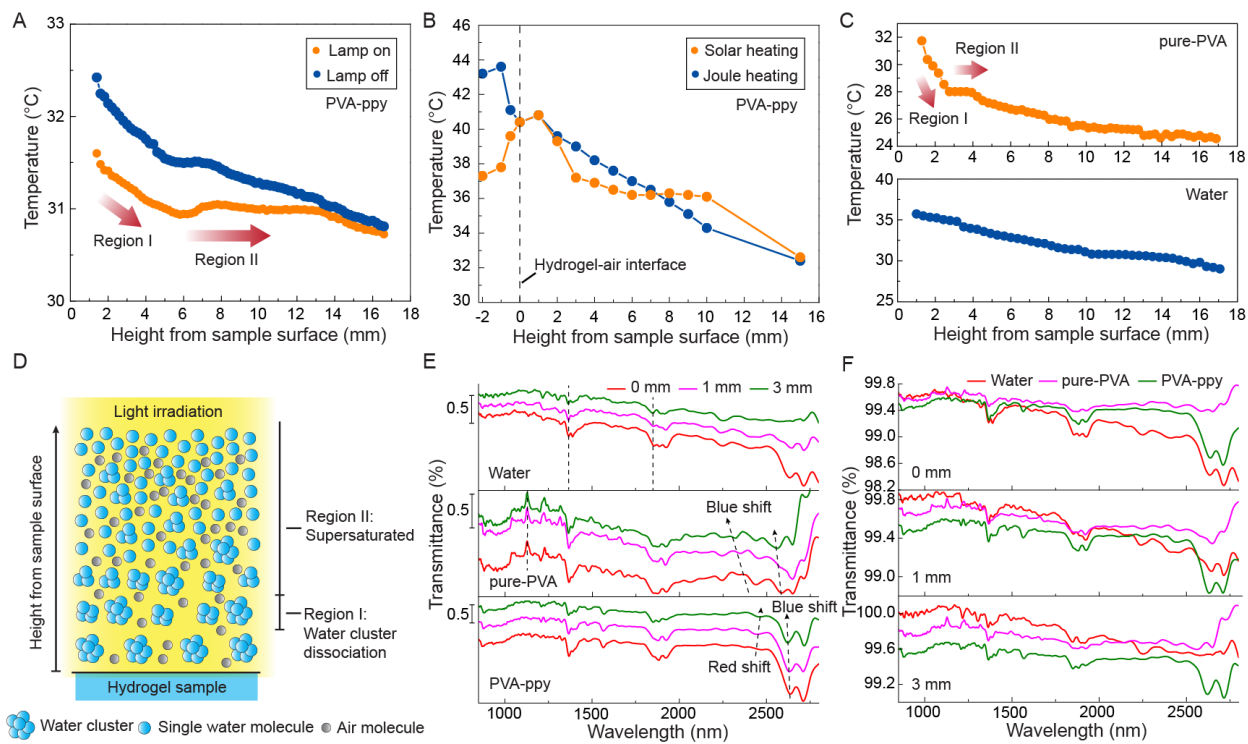

**Fig. 3. Dissociation of water clusters in vapor phase.** (**A**) Vapor phase temperature distributions above PVA-ppy sample measured with IR camera (open dots are surface temperature measured using a thermocouple) when the lamp is on, which shows a sharp temperature drop region (Region I) and a flat temperature region (Region II), and immediately after the lamp is turned off, for which temperature variation is nearly linear. (**B**) Vapor phase temperature distributions of a PVA-ppy sample measured using a thermocouple under solar heating, which is similar to that in 3A with temperature peaks near surface region, and electrical heating, which shows temperature peaks inside and vapor phase temperature distribution similar to lamp off in 3A. (**C**) Comparison of vapor-phase temperature for pure-PVA sample and pure water under green LED (open dots are surface temperature measured using a thermocouple), showing similar behavior as 3A for the PVA-ppy sample under light. The pure water sample has an absorber attached to bottom and sample surface was controlled to be at the same temperature as pure-PVA sample by adjusting the solar intensity. The vapor phase temperature distribution is similar to lamp off in 3A. (**D**) Schematics explaining two different regions regarding dissociation of water clusters. Under light, vapor temperatures drop sharply (Region I) near surface due to dissociation of clusters. A flat region (Region II) exists due to super saturation of water vapor. These features are not seen in thermal evaporation. (**E**) and (**F**) Transmission spectra above water surface at different heights plotted, for different heights but same sample (**E**), and same height but different samples (**F**). In (**E**), spectra are shifted for clarify, and all shifts are within 0.5% of transmittance changes. Absolute transmittance data are shown in (**F**). For pure water, absorption peak does not change. For pure-PVA and PVA-ppy samples, both blue and red shifts are observed progressively away from the sample surface.

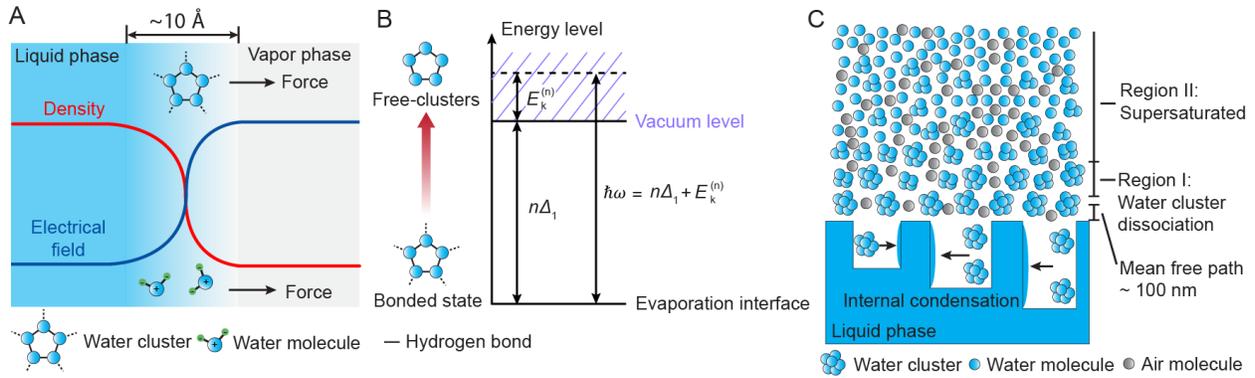

**Fig. 4. Conceptual picture of photomolecular effect.** (**A**) Plausible microscopic mechanism for photomolecular effect at liquid water-vapor interface. The liquid water-vapor interfacial region is ~10 Å. Over similar distance the electrical field also changes rapidly, creating a large electrical field gradient, acting on water clusters. The quadrupole potential generates a large force on the cluster and breaks off the cluster during the cycle when the force points to the vapor phase. (**B**) Energy diagram of the photomolecular effect. A photon with an energy ($\hbar\omega$) larger than the bonding energy ($\Delta_1$ is the average bonding energy per bond and $n$ the number of bonds) between the water cluster and the remaining water liquids can cleave off the water cluster. Excess energy is converted into the kinetic energy of the clusters. (**C**) Interplay between water clusters and structure. After the cluster leaves the interface, it can recondense inside the hydrogel pores (internal photomolecular effect). Water clusters leaving the hydrogel (external photomolecular effect) can be broken up via collision with the air molecules (or among themselves). The breakup region (Region I) is much longer than the air molecule mean free path due to tight hydrogen bonds compared to air molecules' kinetic energy. In this region, heat is absorbed, leading to rapid temperature drop. When the air gets supersaturated, dynamic equilibrium of breakup and recondensation happens (region II), leading to a flat temperature region.

<br>

# Supplementary Materials for

## Photomolecular Effect Leading to Solar-Interfacial Evaporation Above Thermal Limit

Yaodong Tu, Jiawei Zhou, Shaoting Lin, Mohammed AlShrah, Xuanhe Zhao, and Gang Chen

Correspondence to: gchen2@mit.edu; zhaox@mit.edu

**This PDF file includes:**

    Materials and Methods
    Figs. S1 to S14
    Tables S1 to S3
    Movie S1

**Other Supplementary Materials for this manuscript include the following:**

    Data S1: Transmittance spectra above pure-PVA sample at different heights from sample surface under 520 nm LED
    Data S2: Transmittance spectra above PVA-ppy sample at different heights from sample surface under 520 nm LED
    Data S3: Transmittance spectra above pure water at different heights from sample surface under 520 nm LED
    Data S4: Average background spectra without sample nor light used for normalization

**Materials and Methods**

Chemicals

All chemicals, unless specified otherwise, were purchased from Fisher Scientific and used without prior purification. PVA (molecular weight 15K average) was bought from MP Biomedical. APS (≥98%) is acquired from Sigma-Aldrich, pyrrole (reagent grade) is obtained from Aldrich. DI water (18.2MΩ.cm@25ºC) is produced from a Millipore Direct-Q® 5 UV Water Purification System.

Sample preparation

Because the evaporation rate depends sensitively on structures, making samples that can exceed thermal limit requires lots of process optimization. Below, we describe as detail as possible how to make different samples used in our experiments. The process is similar to that described in Ref. (*9*) but also with differences.

**Preparation of PVA aqueous solution.** PVA powder was dissolved in DI water under strong stirring with water bath at 90 ºC for 5 hours to make a 20 wt% solution.

**Preparation of pure-PVA hydrogel samples.** 5 ml PVA solution (20 wt%), 5 ml DI water, and glutaraldehyde (25 wt%) 125 μL were mixed together by a vortex mixer for 2 minutes at room temperature. Bubbles are removed with a centrifuge at 5000 RPM for 3 minutes. Next, HCl aqueous solution (1.2M) 1 ml was added in, mixing with gentle shaking. The obtained solution was injected into petri dish molds to desired thickness and diameter. The gelation was carried out for 2 h at room temperature. The obtained PVA gel was immersed into DI water for 24 hours to obtain pure PVA hydrogel.

**Preparation of freeze-thawed and free-dried pure-PVA samples.** The purified PVA hydrogel was frozen first in a refrigerator at -20 ºC for 6 h. The frozen sample is then moved onto a chilled metal block at -40 ºC (which leads to an initial cooling rate > 5 K/min) for 2 hours. After that, the sample is thawed in open air. To repeat the freeze-thawing, the thawed sample is refrozen on the chilled metal block at -40 ºC for 2 hours and thawed again in open air. To obtain the freeze-thawed sample, the freeze-thawing processes were repeated 1~5 times. The obtained freeze-thawed pure PVA samples are stored inside a humid glass container for future use. To make freeze-dried PVA samples, the above freeze-thawed samples after 3 freeze-thawing is placed on a chilled metal bulk maintained at -60 ºC by liquid nitrogen for 2 hours. Next, the samples are freeze-dried at 0 ºC and 5.2 mTorr in a home-built freeze-drier for 48 hours. The obtained freeze-dried pure PVA samples are stored inside a sealed dry glass container. Fig.S2A&B show photos of the samples.

**Preparation of freeze-dried PVA samples with PPy absorber.** Preparation of PPy aqueous solution, 0.228g of APS particles was dissolved in 20 mL deionized (DI) water. 0.69 mL pyrrole was uniformly mixed with 20 mL DI water by a vortex mixer for 5 min. Adding the APS solution and pyrrole solution dropwise in turn to 50 mL 1.2 M HCl aqueous solution with stir. The polymerization was carried out for 5 min, and quenched with DI water. The as-prepared PPy is purified by filtration and washing using DI water for 3 times. The purified PPy was then well dispersed in DI water by sonication to form PPy solution (20 g/L). Next, PVA aqueous solution (20 wt%) 5 ml, glutaraldehyde (25 wt%) 125 μL, PPy aqueous solution (20 g/L) 5 ml were mixed together by sonication for 10 minutes. Then remove the bubbles with centrifuge machine at 3000 RPM for 5 minutes. Next, HCl aqueous solution (1.2M) 1 ml was added in, mixing with gentle shaking. The obtained solution was injected into petri dish molds. The gelation was carried out for 2 h at room temperature. The obtained PVA-ppy gel was immersed into DI water for 24 hours to obtain the clean PVA-ppy hydrogel. The freeze-thaw and freeze-dry processes are the same as pure PVA samples. The obtained freeze-dried PVA-ppy samples are stored inside a sealed dry glass container. Figs. S2C&D show photos of the samples.

**Preparation of PVA Samples on carbon paper.** Because the original carbon paper (AvCarb MGL190 obtained from Fuel Cell Earth) is a little hydrophobic and its pore size is too large to hold the PVA solution, and its absorptance is relatively low (about 75%), we first treated the carbon paper using a PVA and PPy mixture solution. First, PVA aqueous solution (20 wt%) 1 ml, PPy aqueous solution (concentration 20 g/L) 2 ml were mixed together by a vortex mixer for 2 minutes and then remove the bubbles with a centrifuge at 3000 RPM for 3 minutes. The obtained solution was dropwise and uniformly distributed onto the carbon paper (diameter 35mm, thickness 0.19mm). The wet carbon is dried in open air inside a fume hood. The physically cross-linked PVA will partially fill the big pores in the carbon paper and binder the PPy particles to improve its absorptance up to 95% [Fig.S9(b)]. In the next step, pure PVA layer is coated onto the above-treated carbon paper. PVA aqueous solution (20 wt%) 5 ml, DI water 5 ml, glutaraldehyde (25 wt%) 125 μL were mixed together by a vortex mixer for 2 minutes and then bubbles were removed with a centrifuge at 3000 RPM for 3 minutes. After the carbon paper surface is dry, HCl aqueous solution (1.2M) 1ml was added into the prepared PVA solution, mixing with gentle shaking for 5 minutes. Next, 1 ml the obtained PVA solution is uniformly coated onto the carbon paper. The polymerization was carried out for 2 hours, the purification of as-prepared PVA-carbon sample is the same as pure PVA samples. The obtained PVA-carbon samples are storage inside a water container before testing. Figs.S2E&F show photos of the samples.

Confocal microscopy

Free-thaw pure PVA samples are wet and hence cannot be directly observed inside a SEM. We used confocal microscope (Zeiss LSM 700 Confocal) to observe internal structure of PVA sample. A fluorescein was used as the florescent for the confocal observation by socking the PVA samples in fluorescein solution with a dilution of 1:1000.

Scanning electron microscopy (SEM)

The morphologies of the freeze-dried pure PVA samples and PVA-ppy samples were investigated using scanning electron microscopy (SEM) (Zeiss Sigma 300 VP Field Emission SEM), and the images were analyzed using ImageJ software. The samples were coated with gold before they were placed in the SEM at a low vacuum setting and water vapor was used to decrease the charging effect on them.

Differential scanning calorimetry (DSC)

We measured the average latent heat of water in hydrogel samples by DSC (DSC/cell: RCS1-3277 Cooling System: DSC1-0107). In a typical DSC measurement, we first weighed the total mass of the hydrogel sample. Thereafter, the sample was placed in a Tzero Pan and heated from 30 °C to 200 °C at different heating rates under a nitrogen atmosphere with a flow rate of 50 ml/min (Fig.S3). The curve of heat flow shows a broad peak from 30° to 120°C, corresponding to the evaporation of water in hydrogel. The desorption temperature range depends on the heating rate since the evaporation takes time and the sample may not be at a uniform temperature under fast heating. The integration of the endothermic transition ranging gives the enthalpy for water evaporation. The sample is weighted after measurement to determine the amount of water evaporated. Given the measured water content of the sample, we can calculate the average latent heat of the water in hydrogel samples.

Thermal gravimetric analysis (TGA)

We performed thermal gravimetric analysis (furnace: TGA1-0075, control unit: DCC1-00177) to measure the water content and the decomposition temperature of hydrogel samples. The swollen hydrogels weighing $m_0$ in a titanium pan without any water droplet on the surface of the samples. We first cut a disk-shape of swollen hydrogels of around 10 mg. The swollen hydrogels were heated up from 30 °C to 300 °C at the rate of 5 °C/min. The PVA powder was heated up from 30 °C to 400 °C at the rate of 5 °C/min. All tests were conducted under a nitrogen atmosphere at a flow rate of 30 mL/min. In a typical TGA measurement, the mass of the sample decreases with the increase of temperature and gradually reaches a plateau $m_{dry}$ when all the residual water in the sample evaporates. The water content of the hydrogel sample can be calculated by 1- $m_{dry}/m_0$. Pure-PVA and PVA-ppy samples reach the plateau at lower temperature than PVA powders, because the porous structures accelerate the drying process.

UV-VIS absorptance measurement

We used Cary 5000 UV-VIS spectrometer coupled to an integral sphere (Internal DRA 250) to measure the diffuse reflectance R and transmittance T, and from which the absorptance is obtained by A=1-R-T. For the reflectance measurement, the sample is placed at the back port of the integral sphere (Fig.1C). For transmittance measurement, the sample is placed at the entrance port of the integral sphere. Each measurement is carried in the order of transmittance/reflectance/transmittance, with the third measurement done to make sure that the sample has not changed during the measurements. The reference background spectrum is taken with the backport replaced with a diffuser provided with the integral sphere.

The methods to load the sample onto the integrating sphere are shown in Fig.1C, Fig.S4(A) & (B). Here, the home-made sample holder [Fig.S4(C)] is fabricated from two glass sheets (size 75mmx50mmx1mm) and an acrylic frame (size: outside 75mmx50mmx2mm; inside 69mmx44mmx2mm) as spacer. Noted that the sample fulfilling the cuvette is much larger than the view port of the integrated sphere. The absorption spectrum of the home-made sample holder without samples is shown in Fig.S5(A). Because the sample holder is larger than the entrance port of the integrating sphere, in theory, this kind of sample holder has the risk of overestimating the absorbance because the obtained reflectance is underestimated due to some of the scattering light may escaping from the side wall of the cuvette, which can't be collected by the integrated sphere. To estimate how much this error will be, we also make a similar but smaller cuvette, which can fit into the integrating sphere opening so that all scatterted light can be collected by the integrating sphere. We have checked three different samples with these two holders, including pure DI water (reflection < 5%), as-received PVA powder (reflection > 80%), and freeze-dry PVA sample with 72%wt water (reflection ~ 50%). Based on the measured results, we find the difference is less than 2%.

To clearly show the absorbance of PVA hydrogels, and the water content effects, we also measured the following samples:

1) Glass cuvette: two glass sheets (75mm x 25mm x 1mm) and an acrylic spacer with frame height 2mm and width 3mm. One glass sheet is bond together with the acrylic spacer by super glue, and the other one is removable [Fig.S1(f)].
2) DI water sample: sealed the glass cuvette by super glue and then inject 3 ml DI water into the glass cuvette.
3) PVA solution sample: sealed the glass cuvette by super glue and then inject PVA aqueous solution (10 wt%) 3ml into the glass cuvette.
4) As-gelated PVA sample: put the prepared as-gelated PVA sample (2mm thick) into the cuvette, and then sealed the cuvette with super glue.
5) PVA powder sample: add 2g as-received PVA powder into the cuvette, and then sealed the cuvette with super glue.
6) Freeze-thawed pure-PVA samples with different freeze-thaw cycles: add the prepared freeze-thaw samples into the cuvette, and then sealed the cuvette with super glue.
7) Freeze-dried pure-PVA samples with different water content: put the prepared freeze-dry samples with specified water content into the cuvette, and then sealed the cuvette with super glue.

To control the water contents inside a hydrogel sample, we first measure the weight of a freeze-dried sample, and then transfer the dry sample onto a plain filter paper. Next we spray a mist over the sample with a home-made sprayer in the fume hood. To achieve the specific water content, we repeat spraying for several times. Then we put the wet sample into a sealed petri dish at room temperature for 6 h, and finally weight the wet sample again to obtain the real water content inside the sample. The measured reflectance and transmittance for representative samples are shown in Fig. S6-S7. Figures S5(a) and (b) show negligible absorption in visible range due to glue and in pure water. The 1-2% absorption in water in the visible range can be considered as uncertainly of measurement. Both PVA solution and as-gelated PVA show negligible absorption [Figs.S5(c) and (d)]. Dry PVA powder starts to absorb below 400 nm [Fig.S5e]. Reflectance of freeze-dried samples depends strongly on water content, and has high values at low water content.

Evaporation rate measurement

Figure S8 give some details of evaporation rate measurement platform. An electronic balance (And EK610i) situated on a lift platform is used to monitor the weight loss of the sample during experiments. An air-shield made of transparent plastic film is also located on the lift platform and around the balance to reduce air flow around the sample. The balance is covered with highly reflective mirror film (reflection higher than 90% in visible spectrum measured by Cary 5000 UV-VIS-NIR spectrometer) to minimize heating by the incoming light. The samples with the same size as water container are tightly fixed at the mouth of the water container, where its top surface is at the same height level with the water container. This arrangement avoids the evaporation from the sidewall of the sample on one hand, prevent recondensation that could happen when the sample is below the top edge of the water container on the other hand.

For solar simulator (Sciencetech SS1.4K) driven evaporation, a mirror is used to beam down the solar output (not shown in Fig.S8). The sample stages are carefully insulated using PU foam with about 1 cm in thickness to minimize heat losses. Meanwhile, to avoid the excess light absorption by PU foam, we used aluminum foil to package the PU foam. For LEDs driven evaporation, the sample stage is naked, without any thermal insulation since we found little difference between with and without insulation. To avoid light absorption by the support structure, all the support structure holding the sample stage is made by gluing the glass slides using clear glue (Krazy super glue).

To clearly show the two-stage evaporation, at the very beginning of the measurement, we put a water cap fully covering the hydrogel sample. Evaporation rate without light is recorded and subtracted from the measured evaporation data under light. The two stages of evaporation after light is turned on can be clearly seen from the weight loss as a function of time. One can also clearly observe that when the evaporation rate changes, water recesses into hydrogel.

Here, to make the measurement as accurate as possible, all the tested sample sizes are larger than the inner diameters of the water container about 2 mm. In this case, the samples can be tightly fixed at the mouth of the water container through its friction with the container wall. In this way, we carefully make sure the sample top surfaces are just at the same height with the container mouth. This also only allows the water evaporating from the hydrogel when the water level go below the sample top surface. Normally, we use a laser cutter to tailor the sample size because the as-prepared samples are usually larger than we need. We have also checked the effect of sample size on evaporation rates. For example, PVA-ppy samples under 1 sun, 1cm x 1cm sample's evaporation rate is 3.6 kg/m$^2$h, $\Phi$12mm sample's evaporation rate is 3.2 kg/m$^2$h, however the sample with 25mm in diameter, its evaporation rate is about 2.4 kg/m$^2$h. And more testing results under dark condition are shown in Table S1. Because no rate changes were measured for samples with diameter above 25 mm in diameter, all samples measured have a diameter of 25 mm. The sample holders are made by cutting the glass bottles, with a diameter of 25mm and a depth of 5 mm and wall thickness of 1 mm. The mouth of the container has been carefully polished to make its cross-section flat and smooth. Furthery, the sample holders used for PVA-ppy sample and PVA-carbon sample tests will be well thermally insulated as shown in Fig.S8. However, if they are used for pure PVA sample tests, there is no thermal insulation.

The light intensity is monitored with a thermopile detector (Newport power meter 2936-R with sensor 818P-001-12) to monitor the input light flux intensity before and after one test to make sure the flux intensity is stable. The spatial non-uniformity of the flux intensity of the solar simulator (Sciencetech SS1.4K) is less than 3% for a light spot with 5 cm in diameter. To reduce the noise of the measured weight loss during evaporation, we also carefully check the lab environment to make sure the fluctuation of the signals is no more than 0.01g (balance's precision).

For the wavelength-dependent evaporation measurement, we used LED with different wavelengths. LED lamps were purchased from Chanzon with rated power of 100 W and different wavelengths: purple 390nm, blue 440nm, green 520nm, yellow 590nm, red 650nm, and IR 850nm. To avoid overheating, LEDs need active cooling. We build a water cooling-based LED light source, which not only prevents LED overheating, but also eliminates additional infrared radiation heating, and avoids generating additional air movement which is a very common problem for air cooling. To reduce the divergence and to improve the uniformity of the flux intensity, we built a light guide using highly reflective surface made of a double layer structure: the outside is made of 1 mm aluminum sheet and the inside is a high reflectance plastic sheet mirror (reflectance higher than 92% tested by Cary 5000 UV-VIS spectrometer) glued onto the inside surface of the aluminum cylinder. The LED flux intensity casting onto the sample is measured according to its wavelength.

For the intensity dependence test, the sample stages and all the setups were kept the same. The intensity is varied by changing the input power of the solar simulator (or the LED). For the solar simulator, we measured the average intensity (the average wavelength is 650nm) in 10,000 seconds before and after the test, respectively. For all the tests, the difference is less than 1%. All the wavelength dependence tests and intensity dependence tests are done on the same sample under the same lab environment, but at different times.

To demonstrate purely thermal evaporation does not lead to two-stage evaporation, we used electrical heating to raise the hydrogel's temperature [Fig. S10A]. The main challenge for this approach is that the heating should be nearly uniformly distributed while still allowing water to permeate as in solar irradiation. We tested different ways to heat the sample electrically and eventually settled on hand made mesh heaters. The mesh heaters are made from Nichrome wires 36 Gauge 0.127 mm in diameter, with spacing between wires 0.200 mm. The heaters are embedded into different depths of PVA-ppy hydrogel during the sample preparation. The test configuration is shown in Fig. S10 A&B. We measured thermal evaporate rate for PVA-ppy and PVA-carbon samples by embedding the home-made heaters at different depths from the sample surface. Measurement results are shown in Tables S2 and S3. Thermal evaporation cannot exceed the limit.

We have also tested another way of thermal heating, by solar irradiation from the backside to the bottom of a PVA-carbon sample [Fig. S10C]. In this case, light from the solar simulator is beamed upward by a mirror, going through the bottom of the glass container and water and absorbed by the black carbon and the additional layer of PPy on the black carbon. The light intensity is adjusted to maintain the sample surface at the same temperature as heating from the front surface [Fig. S10D]. FigS10B shows the sample sketch and an IR thermal image. The LED back-side heating shows more uniform temperature than electrical heating.

Temperature measurements

For temperature measurement, we used both thermocouple and IR microscope.

To use IR camera to measure the vapor phase temperature distribution, we use a very thin glass slide (0.1mm in thickness) as a thermal emitter for the IR camera to aim at Fig.S11. The glass slide cannot touch sample surface, so there is a small distance ~0.3 mm, which is added in plotting Fig.3A and 3C, while surface temperatures were measured using thermocouples. The glass is thin and has low thermal conductivity, so that it does not average out the temperature variation in the height direction. Its high emittance favors more accurate temperature measurement. Since our IR camera needs emittance as input, we calibrated the emittance by comparing IR camera reading with thermocouple reading.

To make the IR temperature sensing as accurate as possible, before each IR temperature measurement, we will use a thermocouple to calibrate the IR camara at 0 ºC (insert the glass slide sensor into ice water bath) and 100 ºC (insert the glass slide sensor into boiling water bath), by adjusting its emissivity, reflective temperature, background temperature and the distance between the sample and the lens. The difference at 0 ºC and 100 ºC between IR microscope and thermocouple readings are less than 0.2 ºC and 0.5 ºC, respectively.

Samples of the obtained IR images of sample surfaces are shown in Fig. S11. The line plot of temperature variation along height in Figs. 3(A) and (C) are obtained from averaging the temperature distribution 5 $mm$ around the centerline.

Measuring temperature using thermocouples under light needs special attention. To measure sample surface temperature, thermocouples can be directly embedded underneath the sample surface. However, to measure temperature above the sample in the vapor phase, especially in measuring temperature profile, we took precaution to shield the thermocouple from direct solar irradiation, since poor heat transfer can lead to much higher thermocouple reading due to its absorption of the light. To measure temperature profile in the height direction, it is importance to avoid heat conduction along length of the thermocouple which can average out steep temperature variations in the height direction. We used K-type (Omega fine wire thermocouple CHAL-0005, 125 μm) thermocoupe is shaped into a U-shape (Fig.S12) with horizontal length of 20 mm to minimize heat conduction loss from the junction along the thermocouple wire (*44*), which could also smear out temperature gradient (*44*).

Direct transmission spectrum of vapor phase

To measure the transmission spectrum of the vapor at different height during light-driven hydrogel evaporation operation, a special sample stage was made with water-cooled LED, so that the sample stage can fit into the sample chamber of the UV-VIS-NIR spectrometer (Fig. S13). The probe beam of the spectrometer is limited with an IRIS of the size 1mm (vertical) × 3 mm (horizontal) as shown in Fig. S13. The height of the sample can be controlled with the help of a movable stage. Since the probe beam is fixed, this configuration enables us to scan the direct transmittance of the moist air at different height over the working samples, by lowering the sample. The LEDs are fixed in the UV-VIS sample chamber. We made sure to make the sample at different heights can receive the same input energy (1 sun) by adjusting the LED input voltage. During testing, the sample chamber was kept closed to reduce the noise, but vapor can leak out since the sample chamber is not hermetically sealed. For all these measurements, we always carefully checked but did not find any condensation inside the sample chamber.

For all the tests of PVA-ppy sample and pure PVA sample, the incident light intensity is the same, 1 sun. To heat up DI water under solar radiation, we put a black absorber at the bottom of the sample holder and adjust the incident light intensity to obtain the same evaporation temperature as the PVA-ppy sample at 1 sun. We also carefully measure the direct transmission of dry air. This is used as a common baseline to process the collected data at different heights for various samples. We collect the signals 5 times at each sample height and take their average.

Under LED lamp radiation, the detector of the UV-VIS spectrometer easily gets saturated due to scattered light. To avoid this situation, we put two filters with cutting off wavelength at 530 nm and 610nm in the front of the detector window as shown in Fig. S2 to cut off the lights shorter than these wavelengths.

However, the beam shape coming out of the UV-VIS changes along the pathlength, although long the sample diameter, the change is small. Due to this change, the beam position should be understood as nominal.

We measured transmission spectrum in the vapor phase at different heights on pure-PVA, PVA-ppy surfaces as well as pure water surfaces at different heights, when they are subjected to green light (520 nm). The measured spectra are illustrated in Fig. S14 and also provided as data files since the spectra are rich in information and cannot be easily deciphered at this stage. We have explained some main features which clear show the existences of clusters in the main text. The normalization sometimes leads to over 100% due to environmental drift.

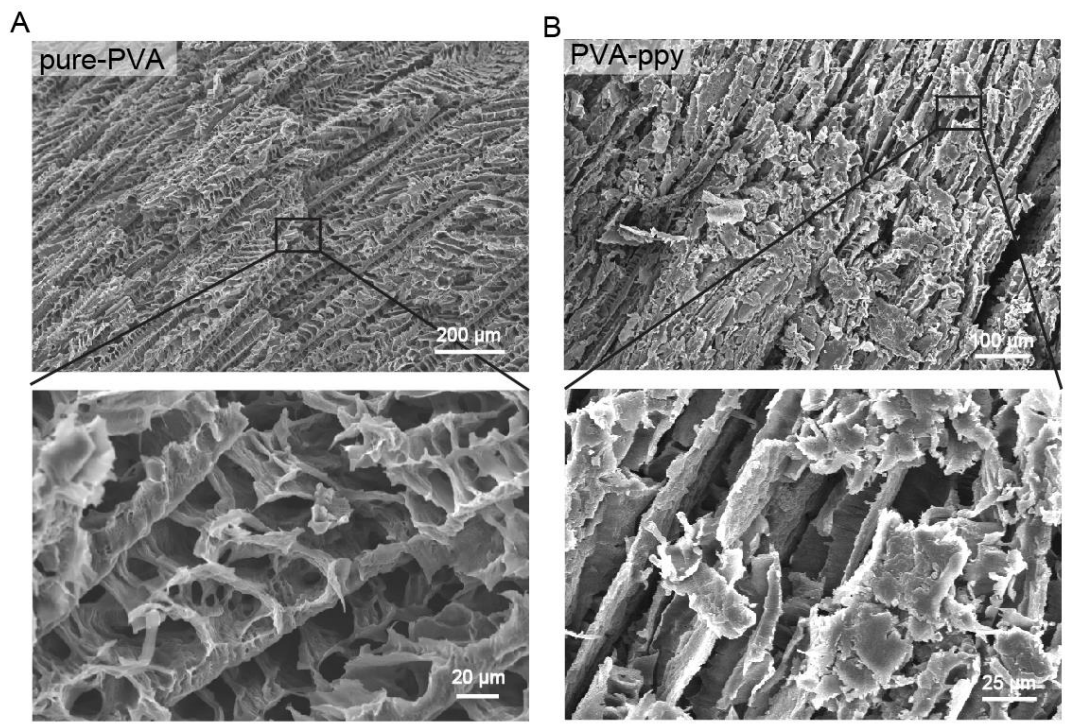

**Fig. S1**. SEM images of **(A)** pure-PVA, and **(B)** PVA-ppy samples. Samples typically show columnar structures running vertically towards surface. The pure-PVA samples have lateral ridges across the vertical columnar structures.

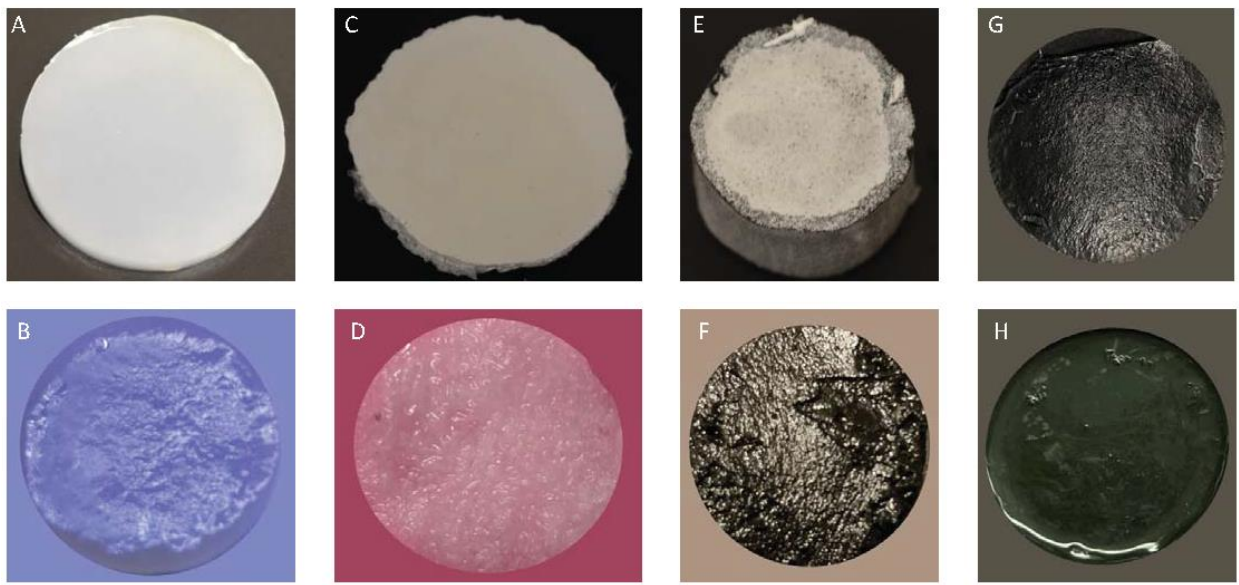

**Fig. S2**. Photos o **(A)** frozen and **(B)** swollen states of freeze-thawed pure PVA sample. **(C)** dry and **(D)** swollen states of freeze-dried pure PVA sample; **(E)** dry and **(F)** swollen states of PVA-PPy sample; **(G)** PPy modified carbon paper at dry state and **(H)** pure PVA coated carbon paper at swollen state.

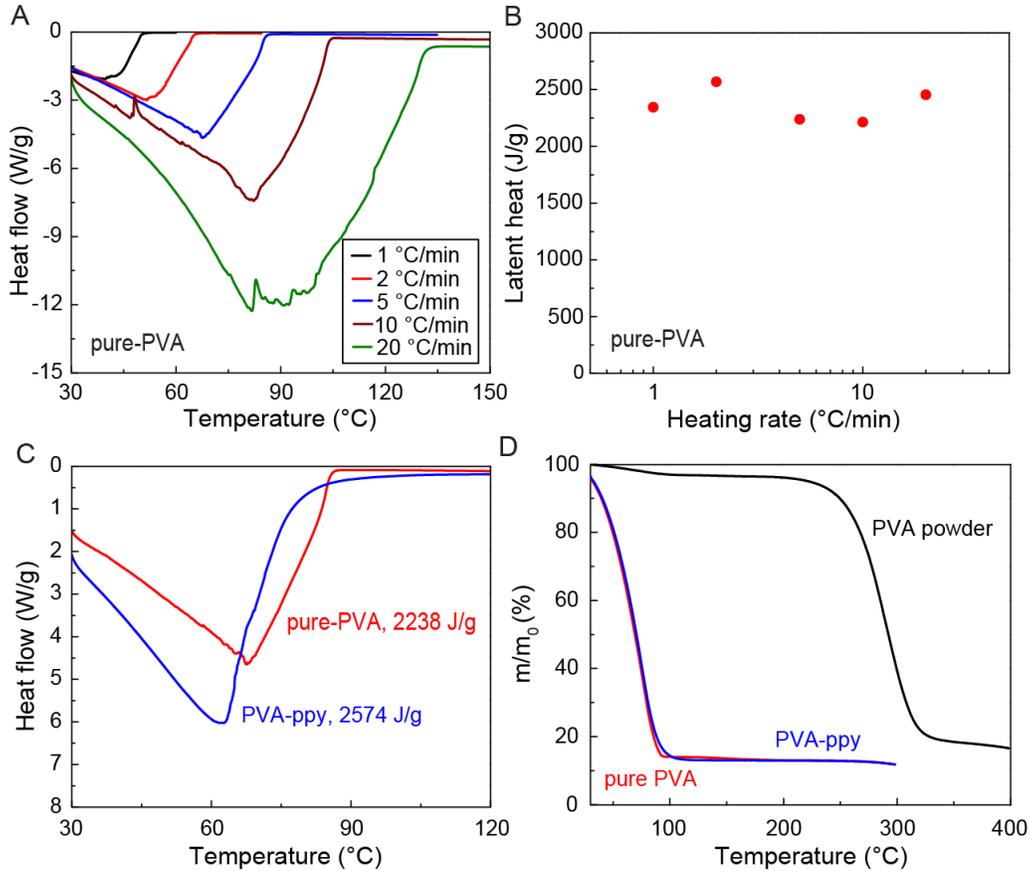

**Fig. S3**. (**A**) DSC thermographs of freeze-dried PVA hydrogels under the scanning rate of 1, 2, 5, 10, and 20 °C/min. (**B**) Measured latent heat of freeze-dried PVA hydrogels versus heating rate. The measured latent heat does not vary much with heating rate. (**C**) DSC thermographs of freeze-dried pure-PVA hydrogel and PVA-ppy hydrogel under the scanning rate of 5 °C/min, giving their average latent heat of evaporation as 2,238 J/g and 2,574 J/g respectively. Pure-PVA samples showed ~13% smaller latent heat than PVA-ppy samples. Such latent heat reduction can be explained with pressure difference between water in hydrogel and pure water at ambient pressure. (**D**) Measurement of the normalized mass reduction (i.e., current mass $m$ over mass at swollen state $m_0$) using TGA of PVA powder, freeze-dried PVA hydrogel, and PVA-ppy hydrogel under the scanning rate of 5 °C/min. The decomposition temperature of PVA-ppy wet sample is around 250 °C obtained from TGA measurement [Fig.S9(c)]. This clearly shows under 1 sun the bonds of the hydrogel matrix will not be broken up with normal evaporation temperature lower than 50 °C. So, the energy to driven evaporation will not come from the matrix decomposition.

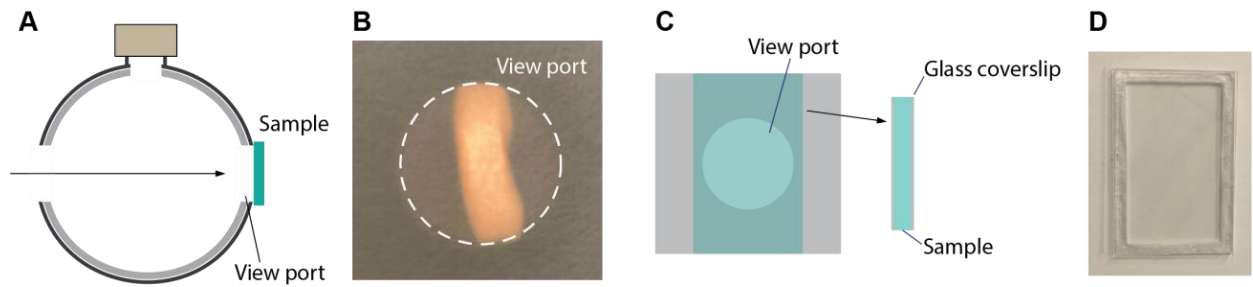

**Fig. S4.** **(A)** Illustration of sample configuration when using the integrating sphere. **(B)** Photo of the back port of integrating sphere and probe beam size. **(C)** Illustration of home-built sample holders for use with integrating sphere, and **(D)** a photo of the sample holder.

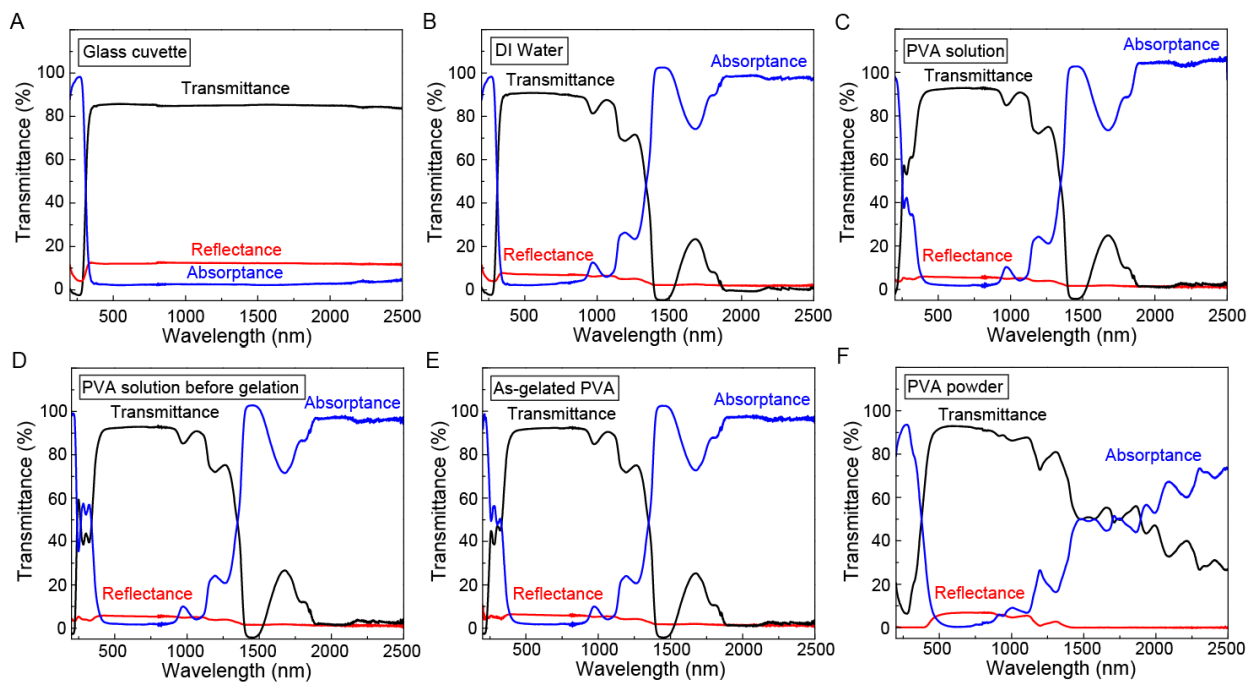

**Fig. S5.** Reflectance, transmittance and absorbance spectra of **(A)** empty glass cuvette; **(B)** DI water; **(C)** PVA solution; **(D)** PVA solution (as-gelated); and **(E)** PVA powder.

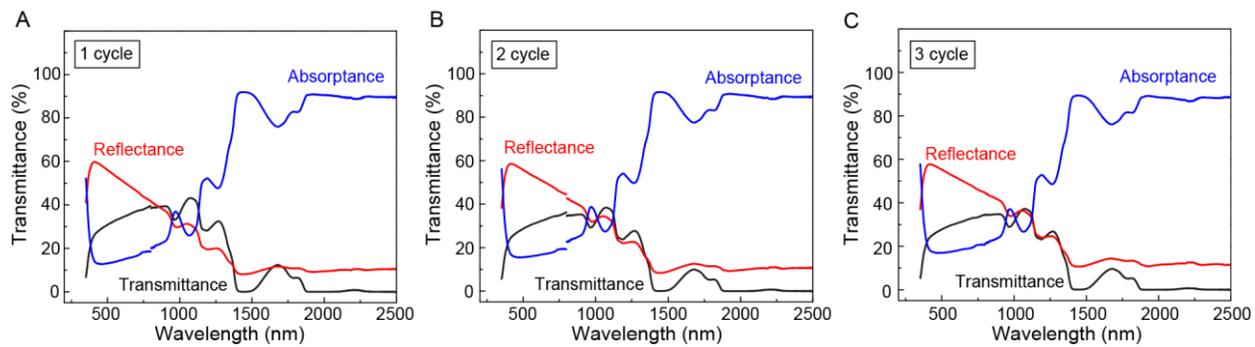

**Fig. S6.** Reflectance, transmittance and absorbance spectra of freeze-thawed samples **(A)** 1 freeze-thaw cycle, **(B)** 2 freeze-thaw cycles, and **(C)** 3 freeze-thaw cycles.

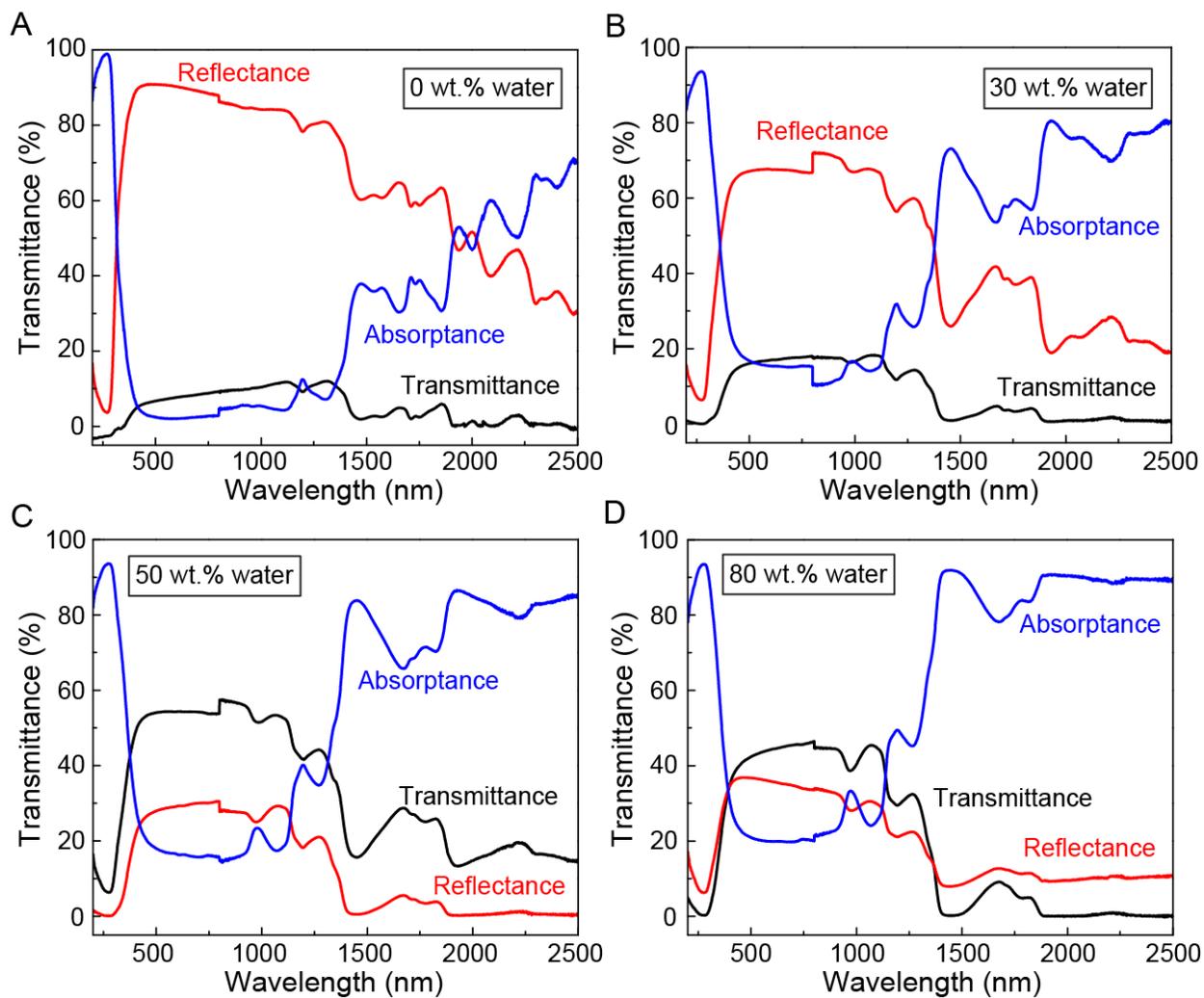

**Fig. S7.** Reflectance, transmittance and absorbance spectra of freeze-dried samples with different water contents: **(A)** 0 wt%, **(B)** 30 wt%, **(C)** 50% wt%, **(D)** 80 wt%. The small discontinuities at 800 nm is due to equipment grating change.

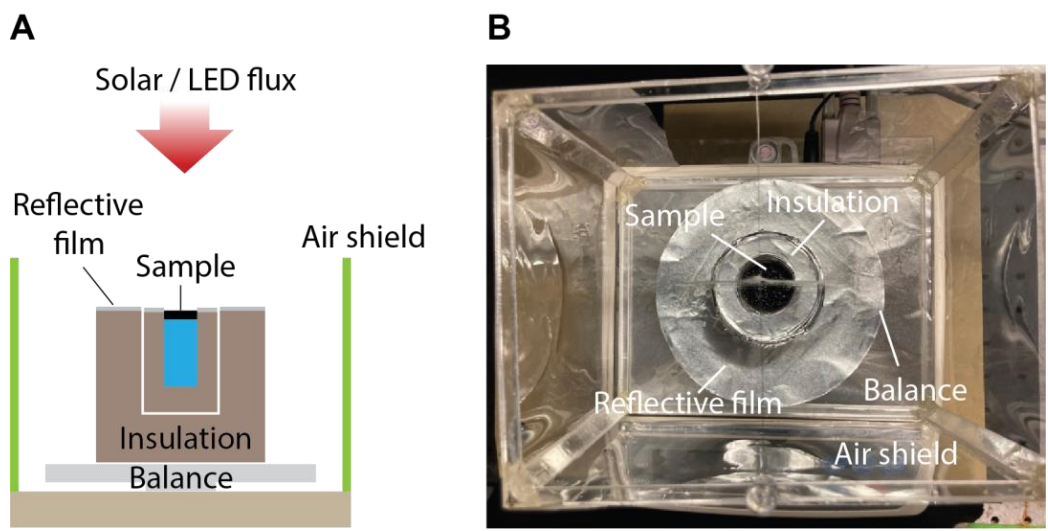

**Fig. S8**. Illustration (**A**) and photo (**B**) of the evaporation rate measurement set up.

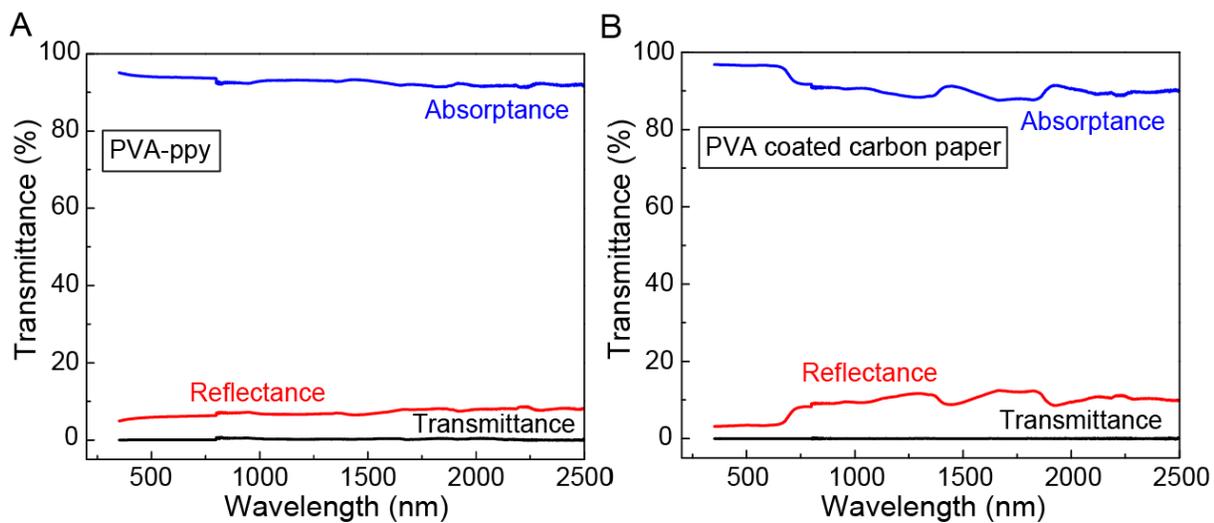

**Fig. S9.** Reflectance, transmittance and absorbance spectra of **(A)** PVA-ppy wet sample; and **(B)** PVA coated carbon paper wet sample.

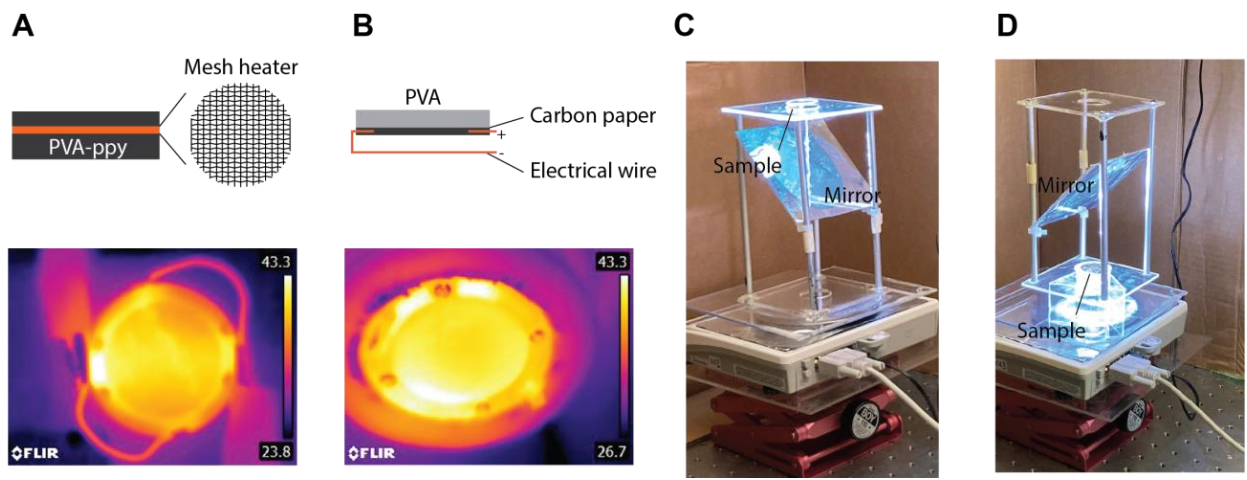

**Fig. S10. (A, B)** Illustration of PVA-ppy and PVA-carbon electrical heating testing configuration and IR image of surface temperature profile, respectively, **(C,D)** photos of a home-made setup that can do solar heating from top and bottom.

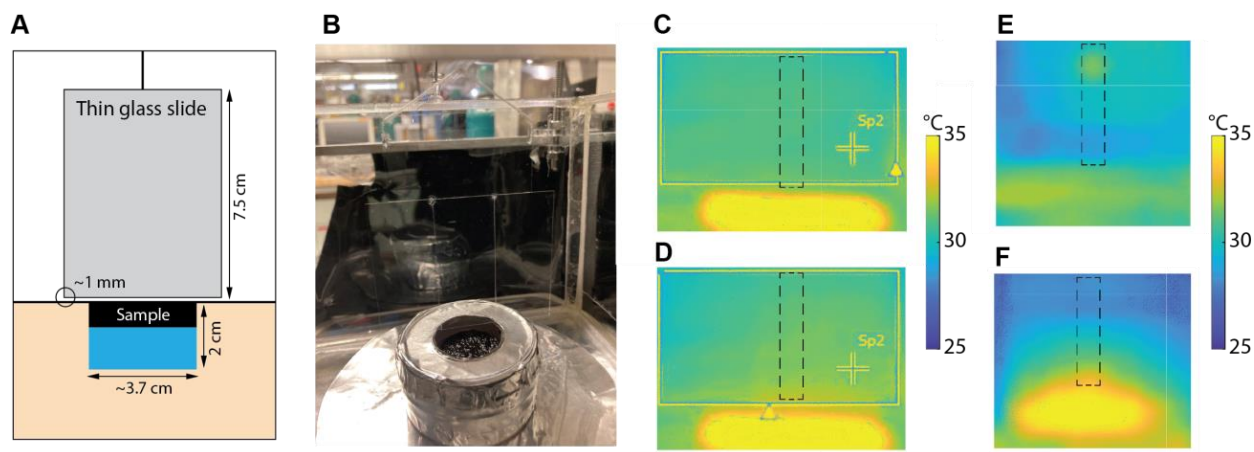

**Fig. S11. Infrared measurement of vapor temperature above evaporating samples.** (**A**) Illustration of thin-glass slide used for IR image of vapor phase temperature; (**B**) Test set up; vapor phase temperature profile for PVA-ppy sample with (**C**) lamp on and (**D**) lamp off; (**E**) vapor phase temperature profile for pure-PVA sample evaporation; (**F**) Vapor phase temperature profile for pure water evaporation. The dashed boxes in (**C-F**) indicate the regions where the temperature profiles are extracted along the vertical direction.

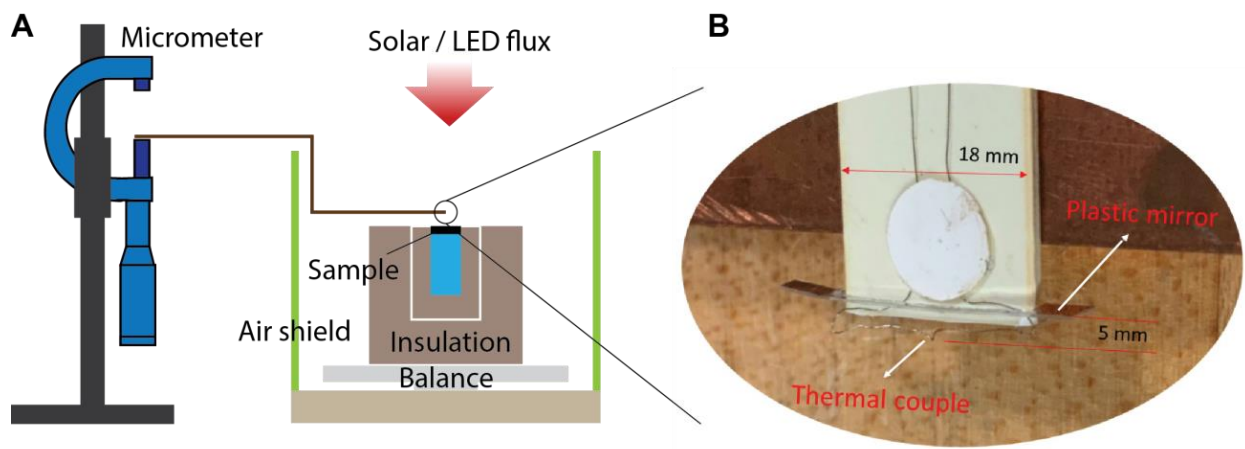

**Fig. S12.** (**A**) Illustration and (**B**) photo of the thermocouple configuration for vapor phase temperature mapping along height direction.

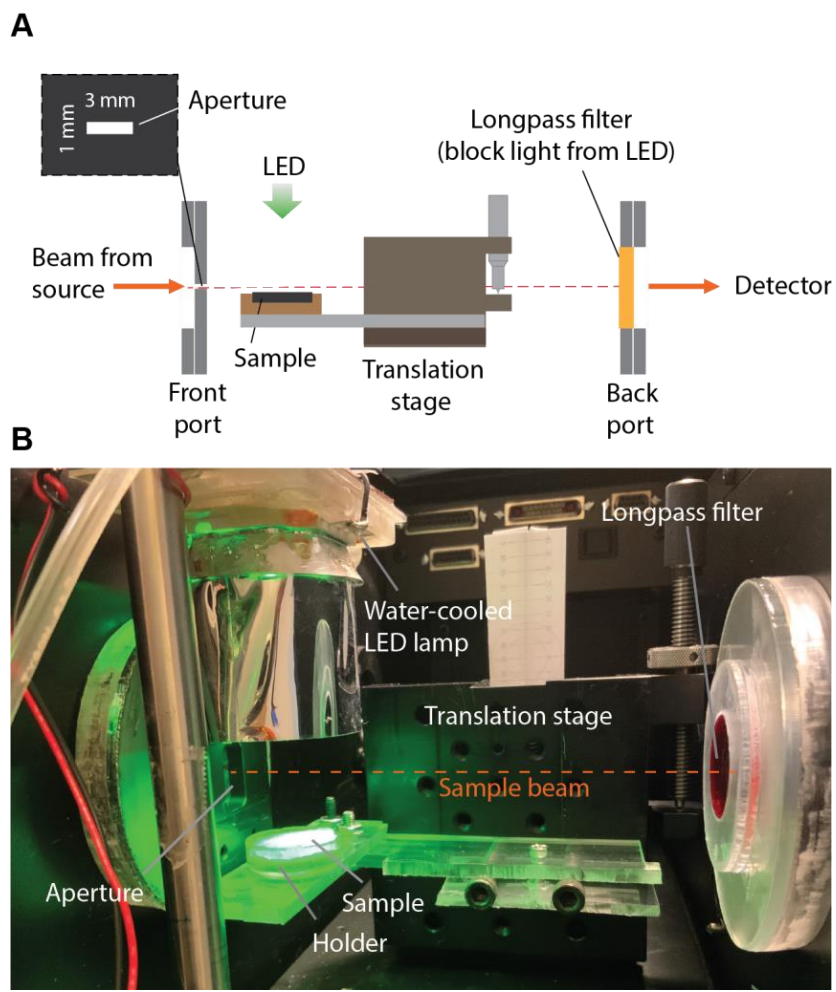

**Fig. S13**. **(A)** Illustration and **(B)** photo of a sample stage for vapor phase absorbance measurement inside a Cary 5000 UV-VIS-NIR spectrometer.

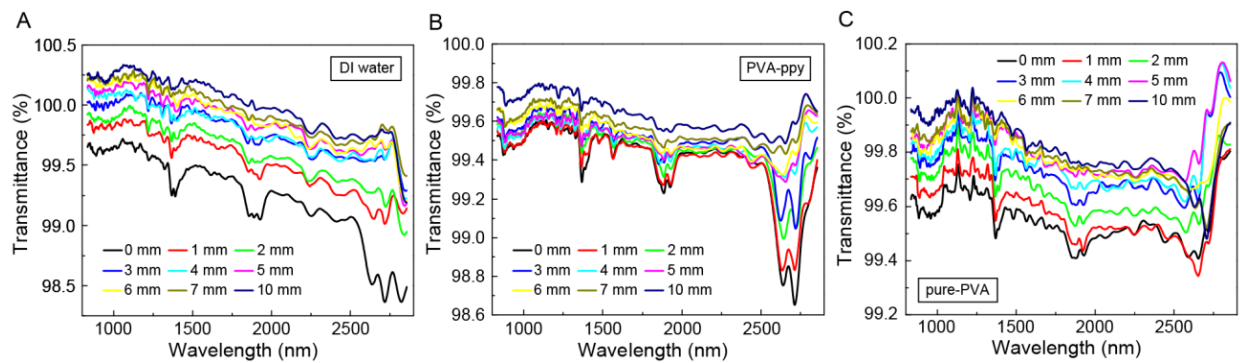

**Fig. S14. Direct transttance spectra of moist air over various samples at different heights**. (A) pure water from 0mm to 10mm; (B) PVA-ppy sample from 0mm to 10mm; (C) Freeze-dried pure PVA sample from 0mm to 10mm. Dry air transmittance, obtained from normalizing the spectra of dry air using an aperture, which limits the spot width in the height direction to ~ 1mm to that without an aperture.

**Table S1. Effect of sample size on evaporation rates under dark condition.**

| Size | PVA-PPy hydrogel sample (kg/m²h) | Pure water (kg/m²h) |
|---|---|---|
| 1 cm × 1 cm | 0.28 | 0.31 |
| Φ 2.5 cm | 0.21 | 0.22 |
| Φ 3.7 cm | 0.21 | 0.22 |
| Φ 6.9 cm | 0.20 | 0.21 |

*All the tests are done at the lab environment, 21.5 ºC / 35% RH, air velocity less than 0.2 m/s. And the water container is the normal petri dish without thermal insulation.

**Table S2 Joule heating efficiency of PVA-PPy sample (4mm in thick) with embedded heating mesh at different height beneath the top surface**\*.

| Position/mm | 4 | 3 | 2 | 1 | 0.5 |
|---|---|---|---|---|---|
| Temperature/oC | 42.8 | 42.5 | 42.6 | 42.9 | 42.5 |
| Evaporation rate/kg/m$^2$h | 0.96 | 1.06 | 1.13 | 1.28 | 1.25 |
| Energy efficiency | 48% | 53% | 79% | 81% | 86% |

\* For the same sample with heating mesh beneath the top surface 1 mm, its evaporation rate is 2.14 kg/m$^2$h @ 1 sun (solar simulator) and its evaporation temperature is 42.5 °C.

**Table S3. Joule heating efficiency of PVA coated carbon paper samples with different coating layer thickness at the same top surface temperature.**

| PVA coating thickness/μm | 300 | 200 | 100 | 40 |
|---|---|---|---|---|
| Evaporation rate @ Joule heating* | 1.29 | 1.36 | 1.39 | 1.43 |
| Joule efficiency** | 83% | 87% | 93% | 96% |
| Evaporation rate @ Solar simulator* | 1.96 | 2.04 | 2.15 | 2.13*** |

\* The evaporation rates have been deducted from the natural evaporation rate (0.22 kg/m$^2$h). The lab environmental condition: 22.4 °C/46%RH. All the testing evaporation temperatures of Joule heating were fixed at the same temperature of 42.5 °C, which is also the evaporation temperature of the same sample under 1 sun (solar simulator).

\*\* The Joule efficiency was estimated by the evaporation rate*latent heat @42.5 °C of water/total input power.

\*\*\*After about 20 minutes of steady evaporation, the sample started to curve up.

**Caption for Movie S1.** Recondensation on a glass slide suspended in the vapor region for PVA-ppy hydrogel under green light (1 sun).